\documentclass[aps,
		prd,
		reprint,
		twocolumn,
		superscriptaddress,
		shortbibliography,
		nofootinbib,
		floatfix,
		notitlepage
		]{revtex4-1}
\pdfoutput=1

\usepackage{amsmath,amssymb}
\usepackage{graphicx,epstopdf}
\usepackage{slashed}
\usepackage{ulem}
\usepackage[dvipsnames]{xcolor}
\usepackage{bbm}
\usepackage{bm}
\usepackage{bbold}
\usepackage{hyperref}
\hypersetup{colorlinks=true}

\newcommand{\be}{\begin{equation}}
\newcommand{\ee}{\end{equation}}
\newcommand{\bea}{\begin{eqnarray}}
\newcommand{\eea}{\end{eqnarray}}

\newcommand{\mbf}[1]{\mathbf{#1}}
\newcommand{\tsf}[1]{\textsf{#1}}
\newcommand{\trm}[1]{\textrm{#1}}

\definecolor{bkcol}{rgb}{0.14,0.42,0.9}
\definecolor{aaaa}{rgb}{0.99, 0.4, 0.01}
\definecolor{bbbb}{rgb}{0.5, 0.3, 0.9}

\newcommand{\ud}{\mathrm{d}}
\newcommand{\LCm}{{\scriptscriptstyle -}} 
\newcommand{\LCp}{{\scriptscriptstyle +}}
\newcommand{\LCpm}{{\scriptscriptstyle \pm}}

\newcommand{\LCperp}{{\scriptscriptstyle \perp}}

\newcommand{\bi}{\begin{itemize}}
\newcommand{\ei}{\end{itemize}}

\makeatletter
\def\ps@pprintTitle{%
 \let\@oddhead\@empty
 \let\@evenhead\@empty
 \def\@oddfoot{}%
 \let\@evenfoot\@oddfoot}
\makeatother

\newcommand{\sfp}{\mathsf{p}}

\newcommand{\sfs}{\mathsf{s}}

\newcommand{\bracket}[2]{\bra{#1}\,#2\rangle} 
\newcommand{\bra}[1]{\langle\,#1\,|}          
\newcommand{\ket}[1]{|\,#1\,\rangle}          
\newcommand{\vphi}{\varphi}
\newcommand{\vtheta}{\vartheta}

\begin{document}

\title{Loop spin effects in intense background fields}

\author{A.~Ilderton}
\email{anton.ilderton@plymouth.ac.uk}
\author{B.~King}
\email{b.king@plymouth.ac.uk}
\author{S.~Tang}
\email{suo.tang@plymouth.ac.uk}

\affiliation{Centre for Mathematical Sciences, University of Plymouth, Plymouth, PL4 8AA, UK}

\begin{abstract}
Radiative and non-radiative electron spin flip probabilities are analysed in both plane wave and focussed laser backgrounds.
We provide a simple and physically transparent description of spin dynamics in plane waves, and demonstrate that there exists a kinematic regime in which the usual leading order perturbative hierarchy of QED is reversed, and non-radiative loop effects dominate over radiative tree-level spin-flips. We show that while this loop-dominance becomes suppressed in focussed laser pulses due to a high sensitivity to field geometry, there is nevertheless a regime in which, in principle, loop effects on spin transitions can be discerned.
\end{abstract}

\maketitle

\section{Introduction}

Intense laser experiments have begun to probe the nonlinear quantum regime of light-matter interactions~\cite{cole18,poder18}. As such, interest has grown in how to properly account for quantum spin effects in strong background fields, in particular within the framework of extended Particle-In-Cell (PIC) codes which underlie the analysis of intense laser experiments~\cite{elkina11,ridgers14,PRE2015Gonoskov}.

Radiative spin-flip in photon emission from an electron in a laser background through nonlinear Compton scattering~\cite{nikishov64,kibble64}, has been studied in a constant crossed field~\cite{ritus85,king15b,Seipt:2020diz} and in pulses~\cite{Boca:2009zz,krajewska13,PRA052117,PRD013010,PRD116001}, for which a comprehensive density matrix formalism has also been developed~\cite{seipt18}. Using the locally constant field approximation (LCFA)~\cite{ritus85}, this radiative spin flip has been included in numerical schemes~\cite{DelSorbo:2017fod,Del_Sorbo_2018,Seipt:2019ddd}. However, the spin can also change in a non-radiative way, through quantum loop effects~\cite{baier76,meuren11,LAVELLE2019135021}. {(See~\cite{Ritus1,Narozhnyi:1979at,Narozhnyi:1980dc,Fedotov:2016afw,Podszus:2018hnz,Ilderton:2019kqp,Ilderton:2019vot,Mironov:2020gbi}, also~\cite{Yakimenko:2018kih,Baumann:2018ovl,Blackburn:2018tsn,DiPiazza:2019vwb}, for investigations of higher loop orders in strong fields, which has recently seen renewed interest.)} It has been suggested to include non-radiative spin changes in numerical codes, using the Bargmann-Michel-Telegdi (BMT) equation~\cite{Li:2019PRL,WAN2020135120,PRAB064401}. However, a generalised BMT equation could in principle include both non-radiative \textit{and} radiative processes~\cite{baier72b}, as it comes from an expectation value. In order to have a consistent approach which avoids e.g.~double-counting, it is important to have a clear understanding of the role of radiative and non-radiative processes in spin dynamics, especially given the known difficulties of closely related topics~\cite{Leader:2013jra}.

In this paper we compare radiative and non-radiative spin flip probabilities in strong background fields. In Sect.~\ref{SECT:SPIN} we first make clear that, in the absence of emissions or loop effects, a plane wave background cannot change the spin state of an electron. From this we clarify the interpretation of the leading order BMT equation in a plane wave background. We then show in Sect.~\ref{SECT:PW} that, for plane waves, even though  the non-radiative loop effect due to spin-flip is $O(\alpha^{2})$ and spin-flip due to radiative effects $O(\alpha)$, there is a kinematic regime where the leading order perturbative hierarchy is reversed, and the loop effect dominates over tree-level effects.

Various mechanisms to extend our results to realistic focussed pulses are discussed in Sect.~\ref{SECT:NEW}. We identify a new shortcoming of the commonly used locally constant field approximation, and analyse the extent to which the `full' BMT equation can be used to model the non-radiative spin flip. This investigation allows us to consider the spin flip in focussed Gaussian pulses in Sect.~\ref{SECT:GAUSS}. We find that while the loop dominance of the plane wave case does not extend to realistic beams, due to focussing effects breaking the symmetry of a plane wave and allowing spin-flipping to occur at the $O(\alpha^0)$ level, show that despite this, the effect of the loop can still be accessed, in principle, through \textit{tail-on} collisions of electrons with laser pulses, in contrast to the commonly considered head-on geometry. We conclude in Sect.~\ref{SECT:CONCS}.

\subsection{Notation and conventions}
Initially we model the laser as a plane wave, thus depending on the lightlike direction $n \cdot x$ where $n^2=~0$. We can always choose $n\cdot x= t-z \equiv x^\LCm$, which is a natural choice of `lightfront' time direction~\cite{Dirac:1949cp,Brodsky:1997de,Heinzl:2000ht,Bakker:2013cea}. The remaining directions are $x^\LCp=t+z$ and $x^\LCperp=(x^1,x^2)$. For momenta we define $p_\LCpm = (p_0 \pm p_3)/2$ and $p_\LCperp=(p_1,p_2)$. The plane wave itself is described by the potential $a_\mu(x^\LCm) = \delta_\mu^\LCperp a_\LCperp(x^\LCm)$ in which the two components of $a_\LCperp$ are the $x^\LCm$--integrals, starting from $x^\LCm=-\infty$ of the wave's two electric field components; this aids the physical interpretation~\cite{Dinu:2012tj}.

\section{Electron spin in laser fields}\label{SECT:SPIN}
Spin in intense laser-matter interactions is commonly considered `classically'~\cite{Li:2019PRL,WAN2020135120} through the BMT equation~\cite{bargmann59} which describes the evolution of the spin vector. However, spin is a purely quantum effect, and we will see that the clearest physical understanding of spin in a laser background ultimately comes from considering the quantum theory. In order to maintain contact with both approaches, though, we will here develop the classical and quantum descriptions somewhat in parallel. 

\subsection{Lightfront helicity} 
An electron has, as well as its momentum $p_\mu$, a spin, which can be expressed in terms of a linear combination of two spin states. For particles \textit{at rest}, these are eigenstates of the spin operator in a chosen direction, with eigenvalues $\pm 1$ (or ``spin up'' and ``spin down''). 
There are many possible choices of basis.

The associated covariant spin vector $s^\mu$ obeys the two conditions $s \cdot p = 0$ and $s \cdot s =-1$. Therefore $s^\mu$ has two degrees of freedom, which corresponds to there being two independent spin states. There are many choices of~$s^\mu$. While all choices agree on what, say, `spin up in the $z$-direction' is for \textit{particles at rest}, there are ambiguities for particles in motion because there are several \textit{inequivalent} ways to boost to the same momentum. 

It is helpful to choose a particular basis of states and corresponding $s^\mu$ which makes calculations simple, and makes the physics manifest both in vacuum and in a plane wave background.  The following simple argument motivates our choice. The only vectors available from which we can construct $s^\mu$ are $p^\mu$ and the propagation direction of the plane wave, $n^\mu$. Taking a linear combination of these, one easily finds that
\be\label{s-sol}
	s^\mu = \pm \frac{1}{m} \bigg(p^\mu - \frac{m^2}{n.p}n^\mu\bigg) \;,
\ee
satisfies the two required conditions $s\cdot p=0$ and $s^2=-1$. For a particle at rest, these vectors are $s^\mu \to \pm(0,0,0,1)$, which is just spin up or spin down in the $z$-direction.

Writing $\sfp \equiv (p^\LCm,p^\LCperp)$ for the three lightfront `spatial' components of momentum, we observe that $\sfs\propto \sfp$, similar to the usual (Jacob-Wick) definition of helicity~\cite{jacob59}, where one would have $\boldsymbol{s} \propto \boldsymbol{p}$ for the Cartesian vector components. This choice of $s^\mu$ is exactly that of `lightfront helicity' used in lightfront field theory~\cite{Brodsky:1997de,Heinzl:2000ht,Chiu:2017ycx}, where $n^\mu$ arises through the choice of time direction. Lightfront helicity states have the special property that the helicity, call it $\sigma$, is equal to the expectation value of the spin in the $z$-direction, $\sigma = \pm 1$, in all Lorentz frames, so that we may talk of spin and helicity interchangeably. For a thorough discussion see~\cite{Chiu:2017ycx}.

The two spin states of the electron are represented by two spinors, $u_{p\sigma }$. Define the Pauli-Lubanski (pseudovector) operator~\cite{Itzykson:1980rh} by 
\be
	{ W}_\mu := -\frac{1}{2}\epsilon_{\mu\nu\zeta\rho} { P}^\nu M^{\zeta\rho} \;,
\ee
where, in the spinor representation, $M^{\zeta\rho} = (i/4)[\gamma^\zeta,\gamma^\rho]$ and ${P}\to p$, the momentum of the state.  The corresponding classical spin vectors are just the expectation values of the Pauli-Lubanski operator in the two spin states: 
\be\label{medel}
	s^\mu =  -\frac2{m} \frac{\bra{p;\sigma } W^\mu \ket{p;\sigma }}{\bracket{p;\sigma}{p;\sigma}}\ \;.
\ee
We want the states corresponding to lightfront helicity. These are the eigenstates of $W_\mu$ contracted with the spin vector~\cite[\S2.2]{Itzykson:1980rh}. Write $h_p^\mu$ for the positive sign solution to (\ref{s-sol}), where subscript $p$ reminds us that the electron 
has momentum $p_\mu$. Defining $L_p \equiv 2h_p \cdot {W}/m$, our states obey
\be\label{eigenstates}
	L_p \, u_{p\sigma} = {\frac{1}{m}} \gamma_5 \slashed{h}_p\slashed{p} \, u_{p\sigma} = \sigma\, u_{p\sigma}  \;.
\ee
Their explicit forms are given in Appendix \ref{Spinform}. In terms of the $u$-spinors,  (\ref{medel}) becomes
\be
	 -\frac1{m^2} \bar{u}_{p \sigma} W^\mu u_{p\sigma} =  \pm h_p^\mu \;. 
\ee
The usefulness of the lightfront helicity basis becomes clear when we turn to the properties of electron spin in an external plane wave.  Here the spin structure of electron states, as described by the Volkov solutions~\cite{volkov35}, changes from $u_{p\sigma}$ to
\be\label{u-pi}
	u_{\pi\sigma}(x^\LCm)  := \bigg( 1 + \frac{\slashed{n}\slashed{a}({x^\LCm})}{2n.p}\bigg) u_{p\sigma} \;.
\ee
To understand the notation $u_{\pi\sigma}$ recall first that the time-dependent \textit{kinetic} momentum of a particle in a plane wave is
\be\label{momentum}
\pi_\mu(x^\LCm) = p_\mu - a_\mu(x^\LCm) + n_\mu \frac{2a(x^\LCm).p-a^2(x^\LCm)}{2n.p} \;,
\ee
and that, by construction, $\slashed{\pi} u_{\pi\sigma}  = m u_{\pi\sigma}$ in order for the Dirac equation to be obeyed. In other words, $u_{\pi\sigma}$ is the $u$-spinor for on-shell momentum $\pi$. Now consider the helicity. A direct calculation shows that the lightfront helicity operator \textit{commutes} with the additional spin structure in~(\ref{u-pi}):
\be\label{dressed-eigenstates}
	L_\pi u_{\pi \sigma} = \bigg( 1 + \frac{\slashed{n}\slashed{a}({x^\LCm})}{2n.p}\bigg)   L_p u_{p \sigma}  = \sigma u_{\pi\sigma} \;.
	\ee
This means that the helicity eigenstates remain eigenstates in the background, even though the momentum of the state changes in time. As this is a basis, we see that the quantum spin state of an electron \textit{cannot be changed by propagation through a plane wave background alone.}

\subsection{The BMT equation}
In light of this result, we turn to the leading order (in $\alpha$) BMT equation. Observe that even though the state of the electron remains unchanged in a plane wave, the associated spin vector does change~\cite{Aleksandrov2020}, from $h_p$ to $h_\pi$. Using (\ref{momentum}) we can express $h_\pi$ in terms of $h_p$ as
\be\begin{split}\label{h-h}
	h^\mu_\pi &\equiv \frac1m\pi^\mu - \frac{m}{n.\pi}n^\mu \\ 
	 &= h_p^\mu - \frac{n \cdot h_p}{n \cdot p}a^\mu + \bigg( \frac{a\! \cdot\! h_p}{n \cdot p} - \frac{a^2 n \cdot h_p}{2n\!\cdot\! p^2}\bigg)n^\mu  \;,
\end{split}
\ee
in which we have used $a \cdot h_p = a \cdot p/m$ and \mbox{$n \cdot h_p = n \cdot p/m$}. Written as in (\ref{h-h}), we recognise $h_\pi$ as nothing but the solution to the leading order BMT equation in a plane wave,
\be\label{10}
	\frac{\ud h_\mu}{\ud x^\LCm} = \frac{1}{n.p} F_{\mu\nu} h^\nu \;,
\ee
with initial condition $h^\mu = h_p^\mu$, where $F_{\mu\nu}=\partial_\mu a_\nu - \partial_\nu a_\mu$ is the electromagnetic field strength tensor. (We have written the equation in terms of lightfront time rather than proper time, as is more common in the literature.) The interpretation of the BMT equation now becomes clear: the classical spin vector precesses, but \textit{only} to account for the fact that the electron momentum is changing, and this is captured by the leading order BMT equation.  It does not describe a change in the spin state of the electron, though, because there is no change. This is shown by the quantum calculations above, and will be recovered from QED scattering calculations below.

\subsection{Summary}
The physics of spin in plane wave backgrounds is made transparent by our calculation: the background changes the momentum of an electron, but not its spin state. We have shown this essential result using a natural lightfront helicity basis, but it can be proven in any basis: for any $s^\mu_\pi$ obeying the lowest order BMT equation, there is a corresponding $L_\pi$ obeying~(\ref{dressed-eigenstates}). Lightfront helicity has the advantages of allowing simple, explicit calculations, including, note, an essentially \textit{algebraic} solution to the leading order BMT \textit{differential} equation, above.

It follows that if one wants to change the spin state of an electron in a plane wave, then emissions or loop corrections are required. This is the topic of the next section, in which we will also see that lightfront helicity \textit{states} have interesting properties.

\section{Spin-flip in plane wave backgrounds}\label{SECT:PW}
The spin state of an electron can be changed in plane wave backgrounds via both radiative and non-radiative processes. Here we describe contributions to each. It is instructive to organise the presentation in powers of the fine structure constant $\alpha$.

Working in the lightfront helicity basis above, the probability of helicity/spin flip from a state $\sigma$ to a state $-\sigma$ in any given process, using $S$-matrix methods and Volkov wavefunctions~\cite{volkov35}, is well-documented~\cite{ritus85,Dinu:PRA052101,king15b,seipt2017volkov,seipt18,baier76,Seipt:2020diz}, so we present only final results here.

\begin{figure*}[t!]
{\includegraphics[width=0.16\textwidth]{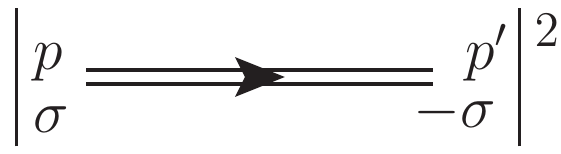}} \quad,\qquad
{\includegraphics[width=0.16\textwidth]{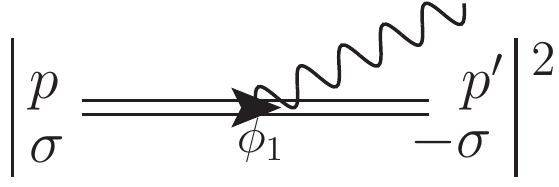}} \quad\qquad
\raisebox{1pt}{\includegraphics[width=0.305\textwidth]{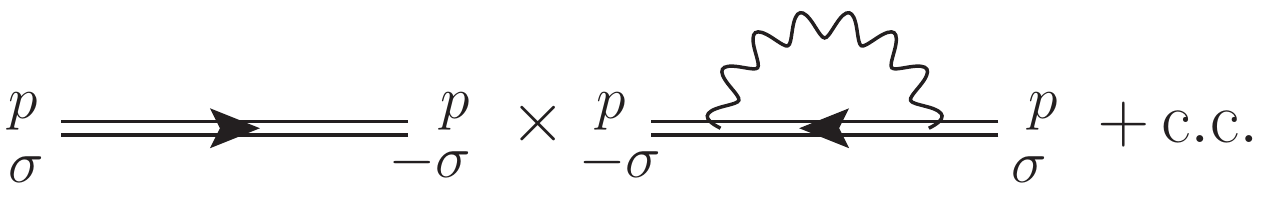}} \quad,\qquad
{\includegraphics[width=0.16\textwidth]{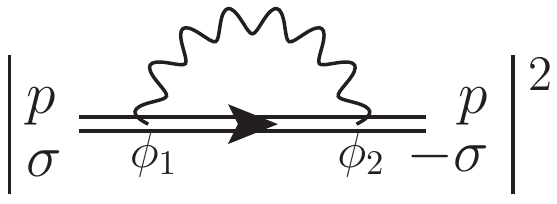}}
\caption{Diagrams contributing to spin flip of an electron, initial momentum $p$ and spin $\sigma$. \textit{Left}: non-radiative spin flip at zeroth order in $\alpha$, from propagation through the field. \textit{Middle two diagrams}: both radiative and non-radiative spin flip receive contributions at order $\alpha$. \textit{Right}: non-radiative spin flip at order $\alpha^2$, from the one-loop propagator correction. Double lines denote background-dressed propagators; diagrams should be understood as LSZ-amputated scattering amplitudes.}
\label{Fig_feynman}
\end{figure*}

\subsection{Order $\alpha^0$: propagation}
%
To zeroth order in $\alpha$, the only contribution to spin flip can be non-radiative, see the left hand diagram in Fig.~\ref{Fig_feynman}. The corresponding scattering amplitude comes from the double LSZ amputation of the Volkov propagator and has the structure~\cite{ilderton2013scattering}
\be
	S_{fi}\propto  {\bar u}_{p\sigma'} \slashed{n} u_{p\sigma} \propto \delta_{{\sigma\sigma'}} \;,
\ee
so that there can be no spin flip. This confirms the result encountered above, that a particle's quantum spin state cannot be changed by propagation through a plane wave alone (i.e.~propagation in the absence of loops or emissions). We underline that it is the contribution from this amplitude which leads to the leading-order BMT equation (\ref{10}), which can then only describe the precession of the classical spin vector, not a change in spin state.

\subsection{Order $\alpha$: radiative spin flip}

The lowest order contribution to radiative spin flip comes at order $\alpha$, from nonlinear Compton scattering (NLC) at tree level. To present the probability,  we define for convenience $\varphi$ and $\vartheta$ as two lightfront times scaled by some typical (central) frequency scale $\omega$ of the plane wave background. The parameter \mbox{$\eta := \omega n\cdot p/m^2$} then characterises the invariant energy of the interaction. Define the floating average \mbox{$\langle f\rangle= \frac{1}{\vtheta}\int^{\varphi+\vtheta/2}_{\varphi-\vtheta/2}d\phi f(\phi)$} for any $f$, and in terms of this the normalised Kibble mass \mbox{$\Lambda=1- \langle a^2 \rangle/m^2+ \langle  a\rangle^2/m^2$}. The probability of (radiative) spin-flip in NLC may then be written
\begin{align}
\tsf{P}_{\tsf{NLC}}= \frac{\alpha}{2\pi\eta}\int\ud\vphi\int^{\infty}_{0}\ud \vtheta~\frac{\partial\textrm{ln}\Lambda}{\partial\vtheta}~\mathrm{C}\bigg(\frac{\vtheta\Lambda}{2\eta}\bigg)\,,
\label{Eq_prob_NLC}
\end{align}
in which the function $\mathrm{C}$ arises as an integral over the lightfront momentum fraction $s$ of the emitted photon, \mbox{$s:=n\cdot \ell/n\cdot p$} for photon momentum $\ell$:
\begin{align}
\textrm{C}(\mu)=\int^{1}_{0}\frac{\ud s~s^2}{1-s}\sin\bigg(\frac{\mu s}{1-s}\bigg)\,.
\label{Eq_S_integral_NLC}
\end{align}
Note that $\tsf{P}_\tsf{NLC}$ is independent of the sign of $\sigma$.

\subsection{Order $\alpha$: interference}
In general, non-radiative spin flip has an order $\alpha$ contribution. This is a quantum interference term, coming from the product of the tree level propagation diagram and the one-loop propagation diagram, see the third diagram in Fig.~\ref{Fig_feynman}.  However, since the order $\alpha^0$ amplitude is diagonal in spin, in a plane wave, the order $\alpha$ contribution is forced to vanish. This result has several important consequences, see below.

\subsection{Order $\alpha^2$: non-radiative spin flip}
\begin{figure}[b!!!]
\center{\includegraphics[width=0.9\columnwidth]{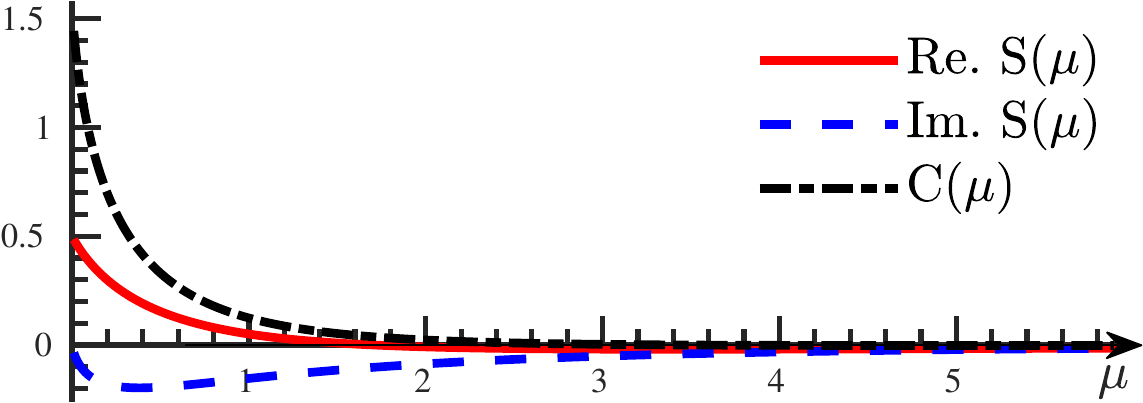}}
\caption{Plots of $\textrm{C}(\mu)$ in (\ref{Eq_S_integral_NLC})
and $\textrm{S}(\mu)$ in (\ref{Eq_S_integral_loop}) showing that these functions do not give significant contributions for large arguments.}
\label{Fig_S_integral}
\end{figure}

Due to the vanishing of the propagation (order $\alpha^0$) and interference (order $\alpha$) terms above, non-radiative spin flip receives its leading order contribution at order $\alpha^2$. This comes from the one-loop correction to the propagator~\cite{baier76,meuren11,Podszus:2018hnz,Ilderton:2019kqp}, mod-squared, see the right hand diagram in Fig.~\ref{Fig_feynman}. With the same notation as used above for nonlinear Compton scattering, the leading order non-radiative spin-flip probability may be written
\begin{align}
\tsf{P}_{\tsf{loop}}=\frac{\alpha^2}{(2\pi\eta)^2} \left|\tsf{M}\right|^2\,,
\label{Eq_prob_loop}
\end{align}
where, for \mbox{$\Delta a_\mu = a_\mu(\varphi+\vartheta/2)-a_\mu(\varphi-\vartheta/2)$},
\be
	\tsf{M} = \int\!\ud\vphi \! \int^{\infty}_{0} \!\frac{\ud\vtheta}{\vtheta}
	\bigg(\frac{\Delta a_{2}}{2m}+i\sigma\frac{\Delta a_{1}}{2m}\bigg)
	\textrm{S}\bigg(\frac{\vtheta\Lambda}{2\eta}\bigg)\,,
\label{Eq_M_loop}
\ee
in which\footnote{We comment that, in the loop calculation, $\varphi$ and $\vartheta$ arise as the average and difference of the lightfront times corresponding to `emission' and `absorption' of the virtual photon, \mbox{$\vtheta=\phi_2-\phi_1$} and $\vphi=(\phi_2+\phi_1)/2$ referring to Fig.~\ref{Fig_feynman}. In the calculation of the NLC probability, if the real photon is emitted at time $\phi_1$ in the scattering amplitude, then $\phi_2$ would be the emission time in the conjugate amplitude.} $\textrm{S}(\mu)$ arises as an integral over the lightfront momentum fraction of the intermediate \textit{virtual} photon,
\begin{align}\label{Eq_S_integral_loop}
	\textrm{S}(\mu) = \int^{1}_{0}\!\ud s~ s~ \mathrm{e}^{-i\mu \frac{s}{1-s}}\,.
\end{align}
With expressions (\ref{Eq_prob_NLC}) and~(\ref{Eq_prob_loop}) in hand we can begin to discuss the physics of the spin-flip probabilities. We first note that the radiative flip probability (\ref{Eq_prob_NLC}) is \textit{independent} of the \textit{sign} of $a_\mu$. The non-radiative flip probability, though, is strongly dependent on the sign of $a_\mu$: because the real/imaginary parts of $\mathrm{S}$ do not change sign (see Fig.~\ref{Fig_S_integral}), the sign of the integrand in (\ref{Eq_M_loop}) is determined by the sign of $a_\mu$. It follows that there can be no non-radiative spin-flip if the potential is an even function, because then the integral over $\vphi$ in (\ref{Eq_M_loop}) gives zero.

This difference in dependence on the driving laser field appears because we look at flips between lightfront helicity states. For electron spins polarised in other directions, the radiative probability can show the same sign dependence as the non-radiative probability, see~\cite{seipt18}.

\subsection{Order $\alpha^2$: radiative}

Radiative spin flip has two order $\alpha^2$ contributions. The first comes from double nonlinear Compton scattering (two photon emission) at tree level, mod-squared. The second is an interference contribution from the cross-term of (single) nonlinear Compton at tree level and at one loop. However, the probability for one spin-flip to occur in double nonlinear Compton scattering scales as $\alpha^2\eta^2\Phi$, where $\Phi$ is the pulse phase duration. We will see that the most interesting region for the loop is when $\eta<\Phi^{-1}$ and when the pulse is short, hence we can ignore order $\alpha^2$ radiative contributions in the regime of interest.

\subsection{Comparison}

The different field dependencies of the spin-flip probabilities found above prompts us to ask when one of the processes, radiative or non-radiative, can dominate over the other. To this end, consider the behaviour of the probabilities for small $\eta$, beginning with the non-radiative result~(\ref{Eq_prob_loop}), which is exact to order $\alpha^2$. The main contribution to the integrand in (\ref{Eq_M_loop}) originates in the region where the argument of $\mathrm{S}$ is small, see Fig.~\ref{Fig_S_integral}. Thus, if \mbox{$\eta\to 0$}, the main contribution comes from $\vtheta\ll 1$. This allows us to approximate the integral, in the small-$\eta$ limit, using a small-$\vartheta$ expansion, setting \mbox{$\Lambda \approx 1-\vtheta^2 a'^2/12\approx 1$}, $\vtheta\Lambda/2\eta \approx \vtheta/2\eta$ and 
$\Delta a \approx m \vtheta (0,\pmb{\varepsilon})$, where $\pmb{\varepsilon}=\pmb{\varepsilon}(\varphi):=\mbf{a}'(\varphi)/m$ is the normalised electric field. Eq.~(\ref{Eq_M_loop}) can then be integrated analytically to give 
\begin{align}
\tsf{M}&\approx \eta \int\ud\vphi  \int^{\infty}_{0} \ud\mu ~\tsf{S}(\mu){\left(\pmb{\varepsilon}_{2} + i \sigma \pmb{\varepsilon}_{1}\right)}\nonumber\\
       &= -\frac{i\eta}{2m}  \big[ a_{2}(\infty)+i\sigma  a_{1}(\infty)\big]\;.
       \label{eqn:16}
\end{align}
Inserting into (\ref{Eq_prob_loop}), the non-radiative spin-flip probability behaves {in the limit $\eta\to 0$ as}
\begin{align}
	\tsf{P}_{\tsf{loop}}\approx \frac{\alpha^2}{(2\pi)^2}\bigg|\frac{{a_{\LCperp}}(\infty)}{2m}\bigg|^2\,,
\label{Eq_Probflip_const}
\end{align}
which is independent of $\eta$. This behaviour holds for pulses with a \textit{unipolar} structure~\cite{Dinu:2012tj,Aleksandrov2020}, $a_{\perp}(\infty)\neq 0$ which essentially means that although the fields oscillate, they are `more' positive than negative (or vice versa). Such pulses exhibit the electromagnetic memory effect~\cite{Dinu:2012tj,Bieri:2013hqa}. For `whole-cycle' pulses, {\mbox{$a_{\perp}(\infty)= 0$}} there is no memory effect, and the leading-order contribution at small $\eta$ is suppressed as a positive power of~$\eta$.

\begin{figure}[t!!!]
\center{\includegraphics[width=1.0\columnwidth]{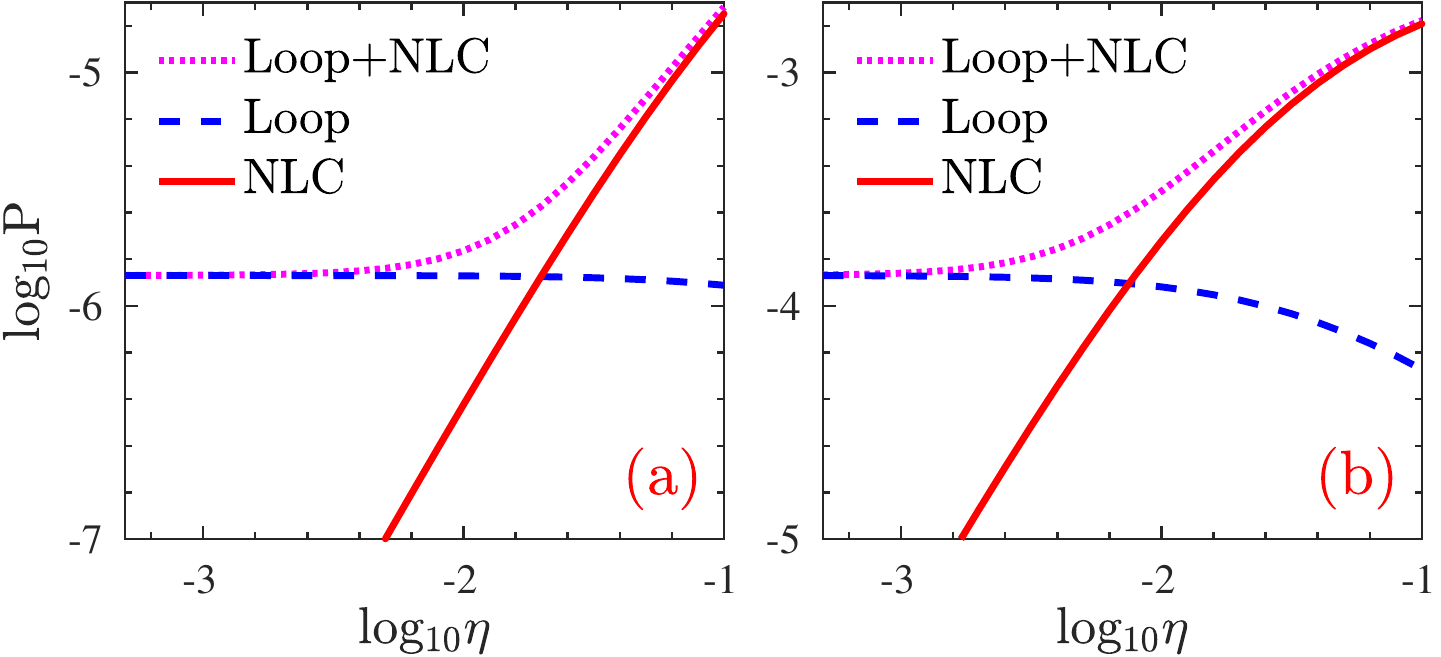}}
\caption{Radiative and non-radiative spin-flip probabilities (\ref{Eq_prob_NLC})  and (\ref{Eq_prob_loop}), respectively, in the head-on collision of an electron with the half-cycle pulse in the text, as a function of varying electron energy from $55$ MeV ($\eta = 10^{-3.3}$)  to \mbox{$11$ GeV} ($\eta=0.1$). Intensities $\xi=1$ (a) and $\xi=10$  (b). The sum of the two probabilities is also shown. The low $\eta$ behaviours (\ref{Eq_Probflip_const}) and (\ref{Eq_NLC_scaling}) are clearly seen.}
\label{FIG:NEW}
\end{figure}

The corresponding low-$\eta$ behaviour of the NLC probability is found similarly; it scales \textit{quadratically} with~$\eta$ as
\be\label{Eq_NLC_scaling}
	\tsf{P}_{\tsf{NLC}}\approx \frac{35 \alpha}{48\sqrt{3}}\, \eta^2\int\ud \vphi \left|\bm\varepsilon(\vphi)\right|^{3}\,.	
\ee
Comparing (\ref{Eq_Probflip_const}) and (\ref{Eq_NLC_scaling}) suggests that at small $\eta$ there can be a reversal of the usual hierarchy of QED perturbation theory, with the one-loop, order $\alpha^2$ non-radiative probability dominating over the tree-level, order $\alpha$, radiative probability.

To demonstrate this result, we consider a head-on collision between an electron and a simple model of a unipolar pulse, being a single half-cycle with electric field ${\bf E}= E_0 (\cos {\omega x^{\LCm}}, 0, 0)$, with \mbox{$|\omega x^{\LCm}|<\pi/2$}, 
meaning normalised potential  \mbox{$a({x^{\LCm}}) =m \xi(0, 1+\sin {\omega x^{\LCm}}, 0, 0)$}, where
\mbox{$\xi := e E_0/(m \omega)$} is the dimensionless peak value of $a$.
We set $\omega = 1.18\,$eV ({$\lambda=1054\,\trm{nm}$}), for an optical laser and vary the electron energy from {$55$ MeV to $110$} GeV, corresponding {to} $\eta$-parameters from $10^{-3.3}$ up to $1$. We plot the exact probabilities (\ref{Eq_prob_NLC}) and (\ref{Eq_prob_loop}) in Fig.~\ref{FIG:NEW}.

We see that for smaller $\eta$, the one-loop non-radiative spin-flip probability is indeed independent of $\eta$, and orders of magnitude larger than the tree-level radiative  probability. The latter increases quickly with $\eta$, confirming the behaviour in (\ref{Eq_NLC_scaling}). The crossover point below which the loop begins to dominate lies around $\eta \approx 0.21\sqrt{\alpha/\xi}$, in which the parameter dependence follows directly from comparing (\ref{Eq_Probflip_const}) and (\ref{Eq_NLC_scaling}). This corresponds to an approximate electron energy of $0.11~(m^2/\omega) \sqrt{\alpha/\xi}$ for head-on collisions. The limits (\ref{Eq_Probflip_const}) and (\ref{Eq_NLC_scaling}) together with the results in Fig.~\ref{FIG:NEW} confirm that loop effects can in principle dominate over tree-level in electron spin dynamics, when the pulse has a unipolar structure. 

\begin{figure}[t!!!!]
\center{\includegraphics[width=0.48\textwidth]{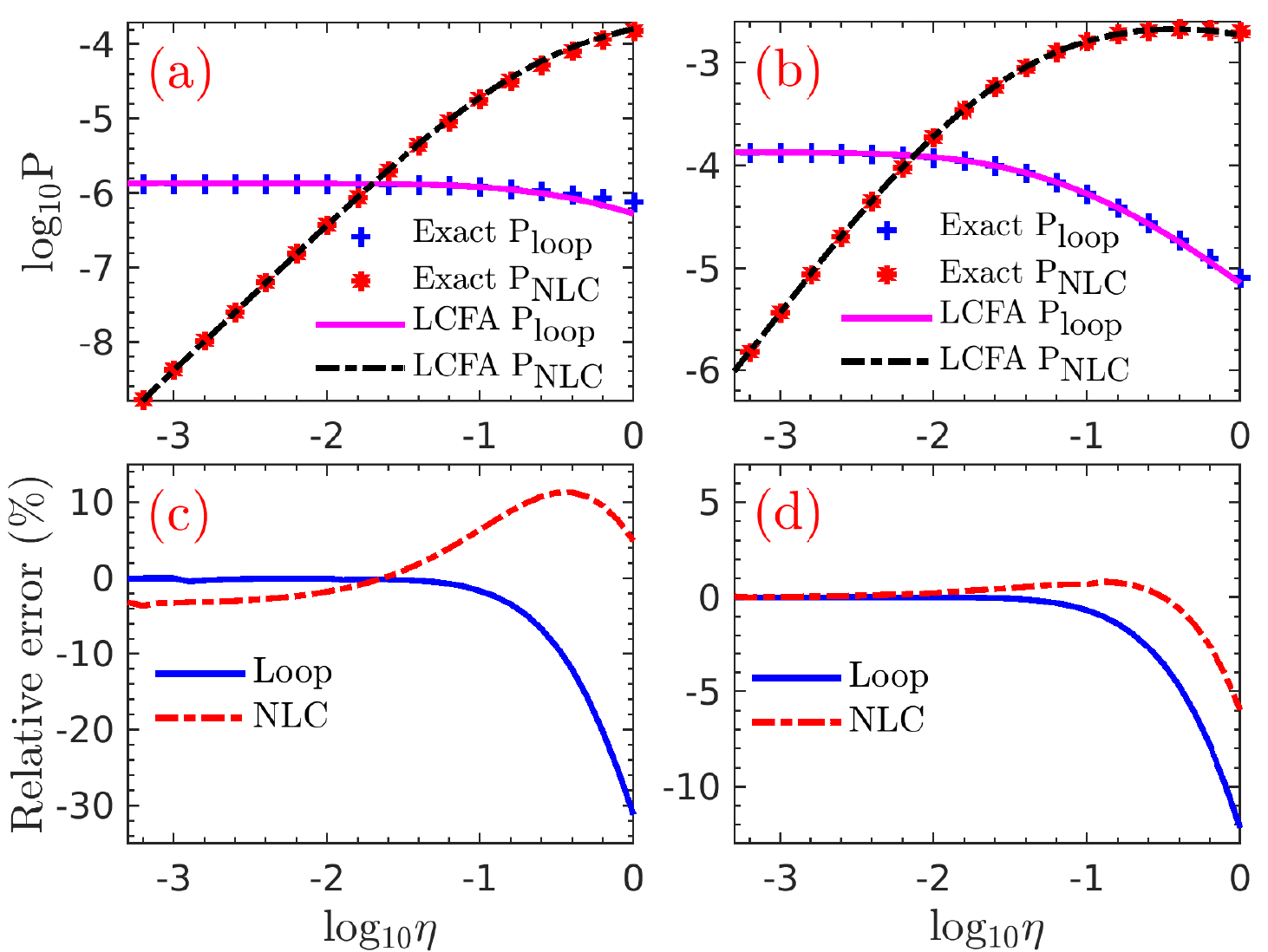}}
\caption{Benchmarking the LCFA  of the radiative and non-radiative spin-flip probabilities against their exact results (\ref{Eq_prob_NLC}) and (\ref{Eq_prob_loop}). {The same parameters as in Fig.~\ref{FIG:NEW} are used.} The upper panels (a) and (b) show the LCFA predictions (\ref{Eq_LCFA}) along with the exact results (\ref{Eq_prob_NLC}) and (\ref{Eq_prob_loop}). The lower panels (c) and (d) show the relative error between the LCFA and exact results, $(\tsf{P}^{\textrm{lcfa}}-\tsf{P}^{\textrm{exact}})/\tsf{P}^{\textrm{exact}}$.}
\label{Fig_Unipolar}
\end{figure}
%

\section{The LCFA and the BMT equation} \label{SECT:NEW}
%
The question we would like to address is whether the loop dominance found above can in principle be observed in realistic electron-laser collisions. The difficulty is that the radiative and non-radiative spin flip probabilities for realistic, strong, fields are not known. The usual approach to sidestepping this problem is to use a `locally constant field approximation' (LCFA) which, applied to the plane wave probabilities, yields approximate expressions which can be extended to more general fields~\cite{ritus85} and implemented in numerical simulations~\cite{PRE2015Gonoskov}. Regarding spin flip, we will identify here some problems with this method, but also present a resolution.

\subsection{Benchmarking}
The standard method to deriving the LCFA from a given plane wave expression is to perform the same small-$\vtheta$ expansion introduced above (\ref{eqn:16}), but now for arbitrary $\eta$~\cite{dipiazza18,PRA2019Ilderton}. This again allows us to integrate over $\vtheta$. The spin-flip probabilities become
\begin{align}
	\label{Eq_LCFA}
	\tsf{P}_{\tsf{NLC}} &= \frac{\alpha}{2\eta}\int\ud\vphi\int^{1}_{0}\ud s \frac{s^2}{1-s}\textrm{Ai}_1(z) \;, \\
	\nonumber
	\tsf{P}_{\tsf{loop}} &= \bigg|\frac{\alpha}{2\eta} \int\!\ud\vphi \! \int^{1}_{0}  \frac{s \ud s}{\sqrt{z}}\left[\textrm{Ai}(z)-i\textrm{Gi}(z)\right]\frac{\varepsilon_{2}+i\sigma\varepsilon_{1}}{|\bm{\varepsilon}|}\bigg|^2\,, 
\end{align}
in which $\textrm{Ai}$ and $\textrm{Gi}$ are the Airy and Scorer functions~\cite{olver2010nist} with argument $z=[s/\chi_p(1-s)]^{2/3}$,  $\chi_p=\eta |\bm{\varepsilon}(\vphi)|$, and $\textrm{Ai}_1(z)=\int^{\infty}_{z}\ud x~\textrm{Ai}(x)$.

The first thing to check is that these approximations are capable of accurately reproducing general plane wave results in the regime of interest. Hence we benchmark the LCFA (\ref{Eq_LCFA}) against exact results in Fig.~\ref{Fig_Unipolar}: they match very well, in particular at small $\eta$ (which justifies the approximation used to find the small $\eta$ limits). In panels (c) and (d) we plot the relative error between the LCFA and exact results; as $\eta$ increases, the relative error becomes larger, and as the intensity $\xi$ increases, the relative error becomes smaller, as expected of the LCFA. We note that the LCFA provides a good approximation even at $\xi=1$.

\subsection{Spin-flip beyond plane waves and BMT}

In general backgrounds, we expect both order $\alpha^0$ (propagation) and order $\alpha^1$ (interference) contributions to non-radiative spin flip. If these contributions are to be approximated using the LCFA applied to plane wave results (or inferred from the constant crossed field result) then they are automatically set to zero. This is another shortcoming of the LCFA, see also~\cite{harvey15,meuren17,Podszus:2018hnz,Ilderton:2019kqp,heinzl2020locally}. As such, we need another method to investigate \textit{non}-radiative spin flip beyond plane waves. One option is to use the full BMT equation for the spin vector~$s^\mu$, but there are subtleties to be confronted in doing so, which we address here.

Let $\ket{\psi}$ be a superposition of single electron lightfront helicity states
\begin{align}\label{tillst}
	\ket{\psi}  = c_\LCp\ket{\pi;+} + c_\LCm\ket{\pi;-} \;,
\end{align}
in which $|c_{\LCp}|^2 + |c_{\LCm}|^2 = 1$ and $\pi$ is the instantaneous (time-dependent) momentum of the state. Using the definition (\ref{medel}) we write the corresponding classical spin vector as 
\begin{align}
	s^{\mu} &= \frac{-2}{m}\bra{\psi} W^\mu \ket{\psi} = \sum_{\sigma',\sigma} c^{*}_{\sigma}c_{\sigma'} S^{\mu}_{\sigma\sigma'}\,,
\label{Eq_BMT_Expansion}
\end{align}
in which the `basis elements' $S^{\mu}_{\sigma\sigma'}$ are given by $S^{\mu}_{\sigma\sigma'} := -2/m \bra{\pi;\sigma} W^{\mu} \ket{\pi;\sigma'}$: for explicit expressions see the Appendix \ref{Spinform}. They obey the relations%
\begin{align}
&S^{2}_{\sigma\sigma'}=-\delta_{\sigma\sigma'}\,,~~S_{-+}\cdot S_{+-}=-2\,,\nonumber\\
&S_{++}\cdot S_{+-}=0\,,~~S_{++}\cdot S_{-+}=0 \;, \nonumber
\label{Eq_spin_basis_relations}
\end{align}
using which it is easily checked that the probability of observing spin up or spin down in the state (\ref{tillst}) can be extracted from the classical spin vector by projecting onto $S_{++}$:
\be\label{find-c}
	|c_\LCpm|^2 = (1 \mp s\cdot S_{++})/2 \;.
\ee
Observe that this result holds in an \textit{arbitrary} background if we take $\pi$ to be the instantaneous momentum in that background, because the lightfront helicity basis is background-independent. As such (noting that the ansatz (\ref{tillst}) explicitly neglects emissions) we have a method for extracting the \textit{non}-radiative spin flip \textit{probability}, in general fields, from the classical spin vector $s^\mu$. This can in turn be calculated using the BMT equation.

Given, then, that we want to use the BMT equation to consider spin beyond plane wave backgrounds, it is worth briefly emphasising what physics the BMT equation does and does not describe. We will continue to focus on plane waves to explore this.
  
The BMT equation as usually used to describe the (non-radiative) evolution of the spin vector is~\cite{bargmann59,baier72b,baier76}:
\be\label{dave}
	\frac{\ud s^{\mu}}{\ud \tau} = \frac{\mu_{b}+1}{m}F^{\mu\nu}s_{\nu}+\frac{\mu_{b}}{m^3}\pi^{\mu}(s_{\alpha}F^{\alpha\beta}\pi_{\beta})\,,
\ee
in which the anomalous magnetic moment {$\mu_b$} is taken to be~\cite{baier72b,baier76,Li:2019PRL,WAN2020135120}
\be\label{mu-def}
	\mu_b = \frac{g-2}{2} = \frac{\alpha}{\chi_p} \int^{1}_{0}  \frac{s \ud s}{\sqrt{z}}\textrm{Gi}(z)\,.
\ee
where $g$ is the electron's gyromagnetic ratio or ``g factor''.
Compare this to the \textit{exact} evolution equation for (\ref{Eq_BMT_Expansion}), which is
\begin{align}
	\frac{\ud s^{\mu}}{{\ud x^{\LCm}}} &={\sum_{\sigma',\sigma}}c^{*}_{\sigma}c_{\sigma'}\frac{\ud S^{\mu}_{\sigma\sigma'}}{{\ud x^{\LCm}}}
	+
	 \sum_{\sigma',\sigma} S^{\mu}_{\sigma\sigma'}\frac{\ud c^{*}_{\sigma}c_{\sigma'}}{{\ud x^{\LCm}}} 
	\nonumber\\
	&=  \frac{1}{n\cdot p} F^{\mu\nu} s_{\nu}+ \sum_{\sigma',\sigma} S^{\mu}_{\sigma\sigma'}\frac{\ud c^{*}_{\sigma}c_{\sigma'}}{{\ud x^{\LCm}}} \;, \label{kate}
\end{align}
where, in the first term, we have used the result~(\ref{h-h}). Using that $\ud x^\LCm / \ud\tau = p^\LCm/m$ in plane waves, we equate (\ref{dave}) and (\ref{kate}), then project onto $S_{++}^{\mu}$ or $S_{+-}^{\mu}$ in order to obtain the evolution equations
\begin{align}
\frac{\ud|c_{-}|^2}{\ud\varphi} &=-\frac{\mu_{b}}{2} \left[c_{-}c^{*}_{+} (\varepsilon_{x}+i\varepsilon_{y})+c_{+}c^{*}_{-} (\varepsilon_{x}-i\varepsilon_{y})\right]\,,\nonumber\\
\frac{\ud c_{-}c^{*}_{+}}{\ud\varphi}&=-\frac{\mu_{b}}{2} \left(1-2|c_{-}|^2\right) (\varepsilon_{x}-i\varepsilon_{y})\,, 
\end{align}
where $\varphi = \omega x^{\LCm}$. For an initial helicity state $\ket{+}$ at $\vphi=-\infty$ we have \mbox{$c_{+}=1$} and $c_{-}=0$. Assuming in the subsequent evolution that $c_{+}\sim 1$ and $c_{-}\sim\alpha$ allows us to approximate the evolution equations as
\begin{align}
\frac{\ud c_{-}}{{\ud\vphi}}=-\frac{\mu_{b}}{2} (\varepsilon_{x}-i\varepsilon_{y}) \;,
\end{align}
and so the spin-flip probability as inferred from BMT becomes
\begin{align}
	|c_{-}|^2=\left|\frac{1}{2}\int \!\ud {\vphi}  (\varepsilon_{x}-i\varepsilon_{y})\mu_{b}\right|^2\,.
\label{Eq_BMT_approximated}
\end{align}
We would expect this to match be the non-radiative spin-flip probability in a plane wave, which we have already seen is well-approximated by Eq.~(\ref{Eq_LCFA}). We see that this matches (\ref{Eq_BMT_approximated}) exactly, \textit{provided} that the Airy function contribution from the loop is neglected. This is negligible, compared to the Scorer function, at small $\chi_p$. Hence, the BMT equation as applied in the literature uses only part of the full loop expression (inferred from the LCFA); it holds only at low $\chi$, which at fixed intensity corresponds to low energy -- this is where the BMT equation can be applied, and this is sufficient for our purposes. Note that in the same limit radiative spin transitions (from photon emission) are suppressed, which is why they can also be dropped from BMT.

\section{Loop dominance in focussed pulses}\label{SECT:GAUSS}
\begin{figure}[t!!!!]
\center{\includegraphics[width=0.48\textwidth]{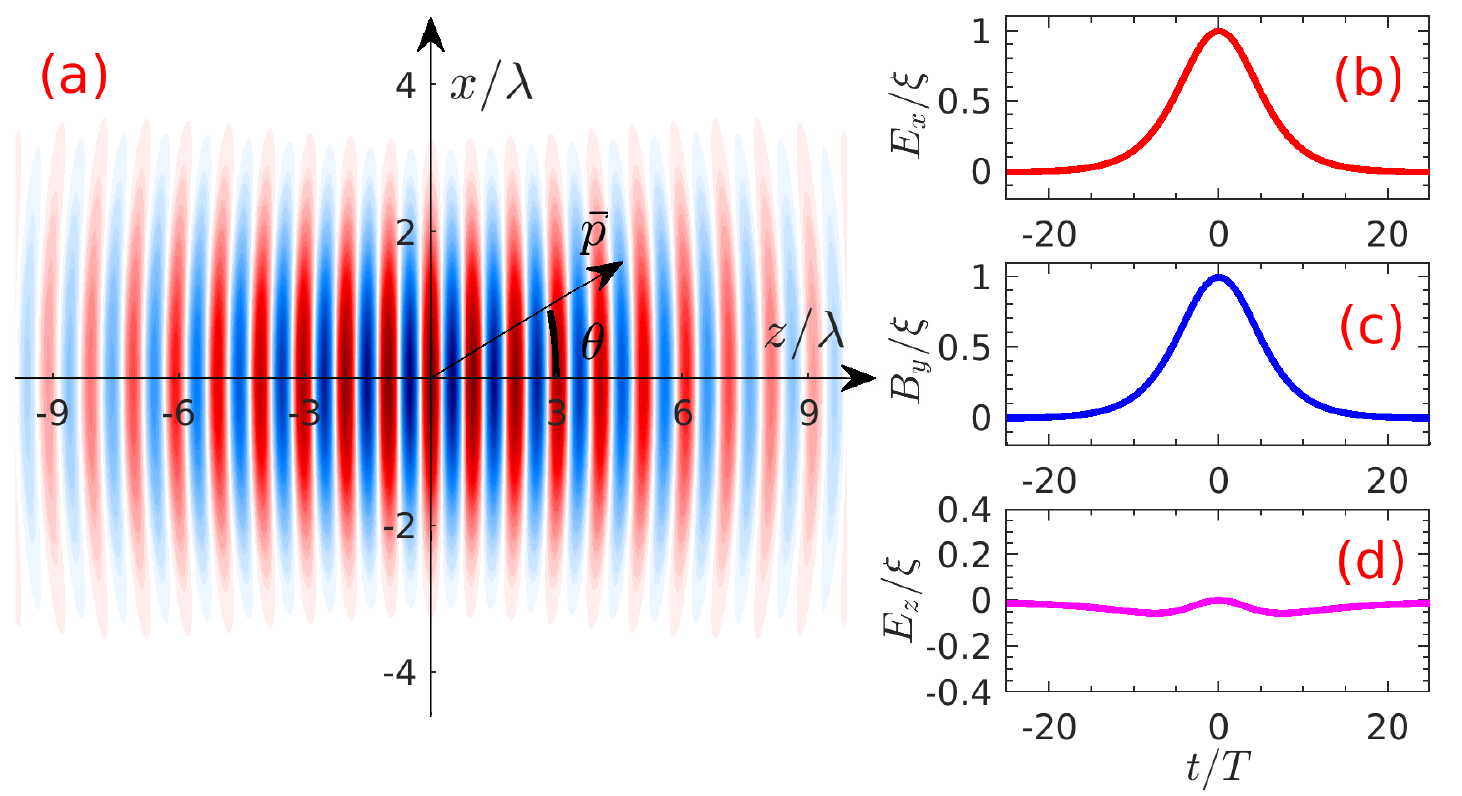}}
\caption{(a) A near \textit{tail-on} collision between a high-energy electron, momentum $p$, and a focussed laser pulse propagating in the $z$-direction, linearly polarised in the $x$-direction. $\lambda$ is the laser wavelength. $\theta$ is the incident angle of the electron in the {$x$--$z$} plane. The electron goes through the pulse centre $(0,0,0)$ at $t=0$. The nonzero components of the electromagnetic fields along the electron trajectory are shown in (b) $E_x$, (c) $B_y$ and (d) $E_z$, in ratio to the peak field amplitude $\xi${: laser focal radius $w=2\lambda$ and electron incident angle $\theta=10^{\circ}$.} } 
\label{Fig_scheme}
\end{figure}
In the above plane wave calculations, a unipolar pulse structure is required to counter the cancellation of opposite-sign contributions in (\ref{Eq_M_loop}), coming from oscillations of the field. What we will now show is that it is possible to realise an \textit{effective} unipolar structure in the collision of electrons with \textit{focussed} laser fields, which in principle allows experimental  access to the loop; we will however see that loop effects are more subtle in focussed pulses than in plane waves.

The key is to consider, in contrast to the de-facto setup of intense laser-matter interactions, an almost \textit{tail-on} collision between the electron and laser. In this geometry a high-energy electron can almost `keep up' with the pulse such that it sees, as it traverses the field, an effectively unipolar, or even same-sign, pulse.  We make this concrete in Fig.~\ref{Fig_scheme}, which shows the effective fields seen by a \textit{high-energy} electron on a ballistic trajectory crossing a tightly-focussed Gaussian pulse. The pulse has focal radius $w$, is linearly polarised in the $x$-direction, propagates in the $z$-direction, and has Gaussian temporal profile $\exp(-t^2/\tau^2)$ where $\tau=5T$ for laser period $T$. (See~\cite{YousefPRL095005} for explicit expressions for the fields.)

Furthermore, in near tail-on collisions $\eta$ is naturally small, \mbox{$\eta=(1-\beta\cos\theta)\gamma\omega/m$} where $\theta$ is the small angular deviation from fully tail-on\footnote{This is provided the electron energy is not too large: this is not a strong constraint as the laser frequency is so low.}. Therefore this interaction setup is that which could allow the dominance of non-radiative loop effects to be seen.

\begin{figure}[t!!!!]
\center{\includegraphics[width=0.96\columnwidth]{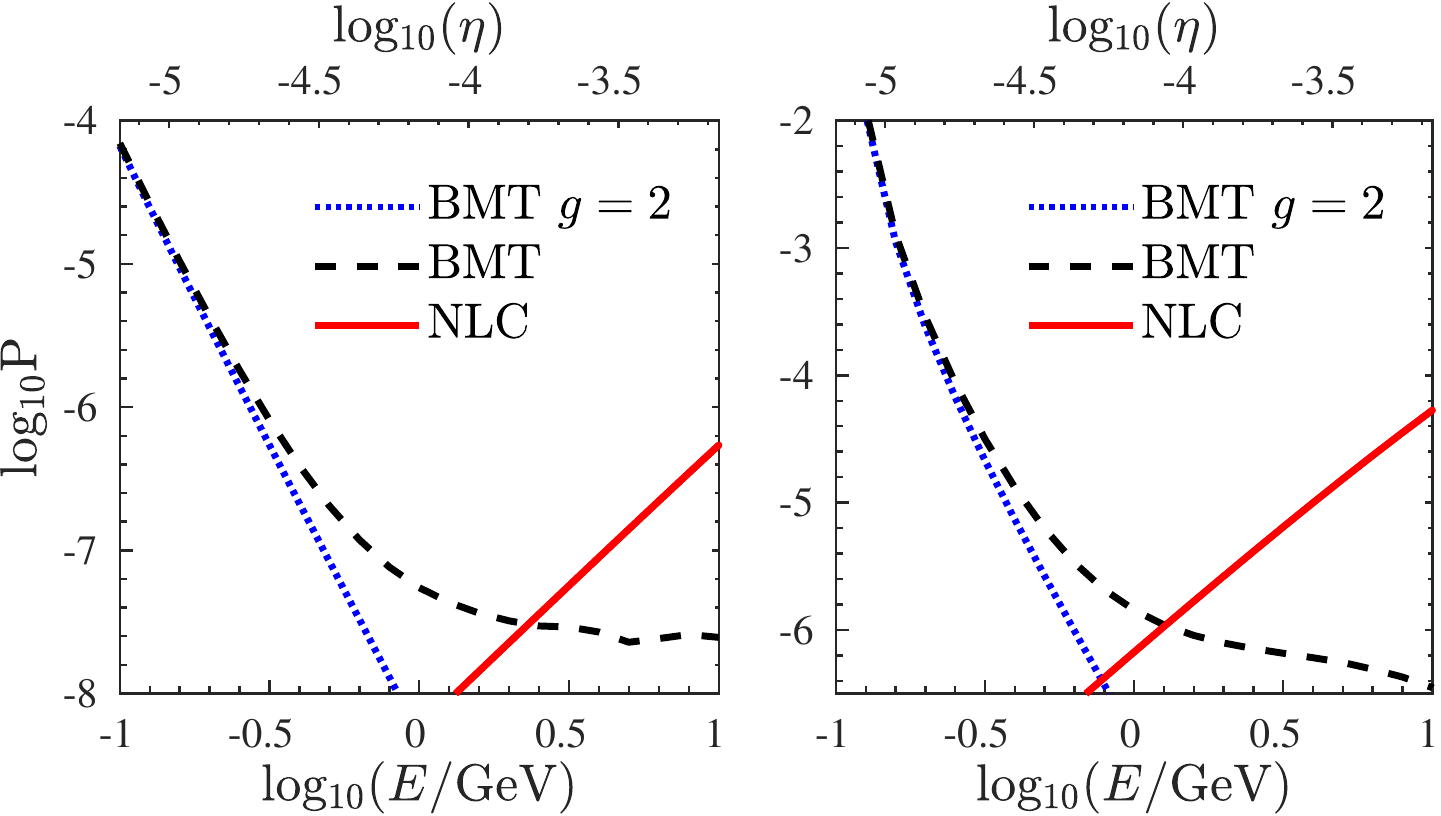}}
\caption{ Spin-flip probabilities in tail-on collisions with \textit{focussed} laser pulses, as a function of initial electron energy, {$w=2\lambda$ and $\theta=10^{\circ}$}. Laser intensity  a) $\xi=10$, b) $\xi=50$. The electron energy changes from $100$ MeV to $10$~GeV. The corresponding change in $\eta$, from $10^{-5.2}$ to $10^{-3.2}$, is shown on the top axis.}
\label{Fig_focus}
\end{figure}

In Fig.~\ref{Fig_focus}, we plot the radiative and non-radiative spin-flip probabilities in near tail-on collisions between a high-energy electron and the Gaussian pulse. The radiative probability is calculated using the LCFA for nonlinear Compton scattering, evaluated along the trajectory of the electron as it traverses the field; this in turn is found by numerically solving the Lorentz force equation. The non-radiative flip probability is calculated by numerically solving the BMT equation (\ref{dave})--(\ref{mu-def}), and extracting from the solution the spin flip probability as in (\ref{find-c}). We present also the non-radiative probability obtained from entirely neglecting the loop, and setting $g=2$ in the BMT equation.

The results do not show the same loop dominance as found in plane waves and as shown in Figs.~\ref{FIG:NEW} and \ref{Fig_Unipolar}. While non-radiative spin flip continues to dominate at low energies, this comes from the leading-order BMT equation in which loop effects are turned off. Our results do show, though, that loop effects are in principle detectable: there is an energy regime, below the region where radiative spin flip contributes, in which the non-radiative spin flip probability is enhanced by the loop by roughly an order of magnitude.

The reason for the difference in behaviour of the spin-flip probabilities in Gaussian beams and plane waves is due to their different field structures, in particular focussing effects.
The symmetry of the plane wave background means that the helicity of the electron cannot be changed at the propagator level, and leading-order spin-flip terms are $O(\alpha^2)$. As soon as this symmetry is broken by focussing, terms of $O(\alpha^0)$ and $O(\alpha^1)$ can contribute to the spin-flip. This breaking of the plane-wave symmetry can be quantified in a focussed Gaussian pulse by the focal expansion parameter $\epsilon :=\lambda/\pi w$ \cite{YousefPRL095005}. In the small-$\eta$ regime, the {leading order} hierarchy of effects contributing to a spin change in a focussed Gaussian pulse is then:
\[
\underbrace{O(\epsilon^{n})}_{\trm{propagator}}\;,
\quad\underbrace{O(\epsilon^n \alpha)}_{\trm{propagator-loop}}\;, \quad\underbrace{O(\alpha^2)}_{\trm{loop}} \;,
\quad\underbrace{O(\eta^2 \alpha)}_{\trm{NLC}} 
\]
For our parameters, as used in Fig.~\ref{Fig_focus}, we have $\epsilon^5\approx 10^{-4}>\alpha^2$ (while higher orders in $\epsilon$ are smaller than $\alpha^2$), so to be consistent in our modelling of the Gaussian pulse, we must use expressions for the field which are correct up to fifth order in the focal expansion parameter. Indeed, in the parameter regime presented in Fig.~\ref{Fig_focus}, numerical analysis using the lower-order beam model suggest that the loop dominance of the plane-wave case persists, but this is misleading, as large differences appear when order $\epsilon^3$ terms are added to the beam, and the loop dominance is mostly lost. However we found that very little change occurred when $\epsilon^5$ terms were added. This is consistent with the order-of-magnitude estimates above. If we had instead used the LCFA approximation (\ref{Eq_LCFA}) and neglected the contribution of lower-order effects from focussing, as is the usual way of applying the LCFA, then only the order $\alpha^2$ spin-flip contribution would have been included. This would have suggested that also in a focussed background, the loop dominance survives, but this would have been incorrect. Thus interference between focussing effects and plane-wave loop effects, is crucial to spin flip at low energy in a laser pulse. We conclude from this that spin-effects described by the BMT equation, in a non-plane-wave background, are very sensitive to the structure of that background. The sensitivity is much higher than can currently be controlled in experiment\footnote{We briefly mention higher-order non-radiative effects. These can be considered using e.g.~the Schwinger-Dyson equations~\cite{meuren11}. In plane waves, this introduces an additional spin structure, with which the helicity operator does not commute -- this is consistent with our results, since loops can change the spin. Such results are still tied to plane waves and hence miss the leading and next-to-leading order terms that are non-zero in a focussed pulse.
}.

\section{Conclusions}\label{SECT:CONCS}
We have analysed radiative and non-radiative electron spin flip in intense fields. In the case of plane wave backgrounds, commonly used as a prototype model of intense laser fields, we have found that there exists a kinematic regime in which the usual leading order perturbative hierarchy of QED is reversed, and where the one-loop non-radiative process dominates over the radiative tree-level process.

This is possible due to two conspiring effects. First, the suppression at low energy (invariant $\eta$) of spin flip due to radiative emissions. Second, the exact vanishing of the \textit{a priori} dominant amplitude for spin-flip through propagation (without emission of loop effects) at tree level.
   
While this loop dominance does not extend to more realistic (e.g. Gaussian beam) models of focussed laser pulses, the influence of the loop on non-radiative spin flip can in principle be accessed experimentally through (near) tail-on collisions between high-energy electrons and tightly-focussed laser pulses. This geometry, in contrast to the usual setup of (near) head-on collisions, suppresses the radiative process, while the Doppler-shifted (\textit{effectively} unipolar) fields seen by the electron on its trajectory enhance the non-radiative processes. Our results thus motivate the study of (near) tail-on collisions of electrons with intense laser pulses, in order to study purely quantum spin effects in intense laser fields.

Our investigation has also underlined several results which have a bearing on other investigations of electron spin in strong fields. First, in plane waves, there can be no spin flip without loops or emissions. As such the leading-order BMT equation commonly studied in the literature describes only the precession of the classical spin vector needed to account for the acceleration of the electron by the field; it does not describe any change in the spin \textit{state} of the electron. Second, because of this result, there is also no order $\alpha^1$ non-radiative spin flip in a plane wave, and so any attempt to infer from plane wave results (e.g.~via the LCFA) for more general, focussed beams, fail outright.  (This is in contrast to radiative spin flip through nonlinear Compton.) As we saw above, predictions which neglect the order $\alpha$ interference terms are incorrect.

Third, in focussed laser pulses, focussing effects dramatically alter the dynamics of non-radiative spin-flip compared to the plane wave case. The sensitivity of spin flip to focussing effects in the background can be greater than that to loop effects, meaning that focussed fields must be modelled to very high precision in order to correctly report spin-flip effects where non-radiative processes play a significant role.  (This is at low energy, since NLC dominates spin flips at higher energies.) This would suggest that extreme care is needed, in simulations and experimental modelling, in parameter regions where BMT contributes an appreciable amount to the overall probability for flip of the electron spin. 

\subsection*{Acknowledgments}
 The authors are supported by the EPSRC, Grant No.~EP/S010319/1.

\appendix

\section{Spin basis}\label{Spinform}
Let ${\sigma}_{i}$ be the $2\times 2$ Pauli matrices. In the Kogut-Soper basis for the $\gamma$-matrices~\cite{Brodsky:1997de},
\begin{align}
\gamma^{0}&=\begin{bmatrix}
    0 & \mathbb{1}  \\
    \mathbb{1}  & 0
\end{bmatrix}\,,~
\gamma^{1}=\begin{bmatrix}
    0 & -\sigma_1 \\
   \sigma_1 & 0
\end{bmatrix}\,,~\nonumber\\
\gamma^{2}&=\begin{bmatrix}
    0 & -\sigma_2 \\
    \sigma_2 & 0
\end{bmatrix}\,,~
\gamma^{3}=\begin{bmatrix}
    0 & -\sigma_3 \\
   \sigma_3 & 0
\end{bmatrix}\,,\nonumber
\end{align}
and $\gamma^5=i\gamma^0\gamma^1\gamma^2\gamma^3$, the explicit form of the lightfront helicity spinors $u_{p\sigma}$ in our discussion is  
\begin{align}
u_{p+}&=\frac{1}{\sqrt{2mp^{-}}}
\begin{bmatrix}
    p^{1}-ip^{2}\\
    p^{-}\\
    0 \\
    m
\end{bmatrix}\,,\nonumber\\
u_{p-}&=\frac{1}{\sqrt{2mp^{-}}}
\begin{bmatrix}
    m\\
    0\\
    p^{-}\\
  -p^{1}-ip^{2}\\
\end{bmatrix}\,.\nonumber
\end{align}

In cartesian components, the basis elements used in the BMT calculation, as defined in and below (\ref{Eq_BMT_Expansion}), obey
\begin{subequations}
\begin{align}
S^{\mu}_{++} &= -S^{\mu}_{--} = h^{\mu}_{\pi}\,,\\
S^{\mu}_{+-} &=~\bigg(\frac{\pi^{1}+ i \pi^2}{\pi^{\LCm}}\,,1\,,i\,,\frac{\pi^{1}+i \pi^2}{\pi^{\LCm}}\bigg) \;, \\
S^{\mu}_{-+} &= S^{\mu*}_{+-}  \;.
\end{align}
\end{subequations}
{where $n\cdot S_{++}=\pi^{-}/m$ and $n\cdot S_{+-}=0$.}
Note that $\pi^\mu$ here stands for the in-principle time-dependent momentum of the state.


\begin{thebibliography}{61}%
\makeatletter
\providecommand \@ifxundefined [1]{%
 \@ifx{#1\undefined}
}%
\providecommand \@ifnum [1]{%
 \ifnum #1\expandafter \@firstoftwo
 \else \expandafter \@secondoftwo
 \fi
}%
\providecommand \@ifx [1]{%
 \ifx #1\expandafter \@firstoftwo
 \else \expandafter \@secondoftwo
 \fi
}%
\providecommand \natexlab [1]{#1}%
\providecommand \enquote  [1]{``#1''}%
\providecommand \bibnamefont  [1]{#1}%
\providecommand \bibfnamefont [1]{#1}%
\providecommand \citenamefont [1]{#1}%
\providecommand \href@noop [0]{\@secondoftwo}%
\providecommand \href [0]{\begingroup \@sanitize@url \@href}%
\providecommand \@href[1]{\@@startlink{#1}\@@href}%
\providecommand \@@href[1]{\endgroup#1\@@endlink}%
\providecommand \@sanitize@url [0]{\catcode `\\12\catcode `\$12\catcode
  `\&12\catcode `\#12\catcode `\^12\catcode `\_12\catcode `\%12\relax}%
\providecommand \@@startlink[1]{}%
\providecommand \@@endlink[0]{}%
\providecommand \url  [0]{\begingroup\@sanitize@url \@url }%
\providecommand \@url [1]{\endgroup\@href {#1}{\urlprefix }}%
\providecommand \urlprefix  [0]{URL }%
\providecommand \Eprint [0]{\href }%
\providecommand \doibase [0]{http://dx.doi.org/}%
\providecommand \selectlanguage [0]{\@gobble}%
\providecommand \bibinfo  [0]{\@secondoftwo}%
\providecommand \bibfield  [0]{\@secondoftwo}%
\providecommand \translation [1]{[#1]}%
\providecommand \BibitemOpen [0]{}%
\providecommand \bibitemStop [0]{}%
\providecommand \bibitemNoStop [0]{.\EOS\space}%
\providecommand \EOS [0]{\spacefactor3000\relax}%
\providecommand \BibitemShut  [1]{\csname bibitem#1\endcsname}%
\let\auto@bib@innerbib\@empty
\bibitem [{\citenamefont {{J.M.~Cole et al}}(2018)}]{cole18}%
  \BibitemOpen
  \bibfield  {author} {\bibinfo {author} {\bibnamefont {{J.M.~Cole et al}}},\
  }\href {\doibase 10.1103/PhysRevX.8.011020} {\bibfield  {journal} {\bibinfo
  {journal} {Phys. Rev. X}\ }\textbf {\bibinfo {volume} {8}},\ \bibinfo {pages}
  {011020} (\bibinfo {year} {2018})}\BibitemShut {NoStop}%
\bibitem [{\citenamefont {{K.~Poder et al}}(2018)}]{poder18}%
  \BibitemOpen
  \bibfield  {author} {\bibinfo {author} {\bibnamefont {{K.~Poder et al}}},\
  }\href {\doibase 10.1103/PhysRevX.8.031004} {\bibfield  {journal} {\bibinfo
  {journal} {Phys. Rev. X}\ }\textbf {\bibinfo {volume} {8}},\ \bibinfo {pages}
  {031004} (\bibinfo {year} {2018})}\BibitemShut {NoStop}%
\bibitem [{\citenamefont {Elkina}\ \emph {et~al.}(2011)\citenamefont {Elkina}
  \emph {et~al.}}]{elkina11}%
  \BibitemOpen
  \bibfield  {author} {\bibinfo {author} {\bibfnamefont {N.~V.}\ \bibnamefont
  {Elkina}} \emph {et~al.},\ }\href@noop {} {\bibfield  {journal} {\bibinfo
  {journal} {Phys. Rev. ST Accel. Beams}\ }\textbf {\bibinfo {volume} {14}},\
  \bibinfo {pages} {054401} (\bibinfo {year} {2011})}\BibitemShut {NoStop}%
\bibitem [{\citenamefont {Ridgers}\ \emph {et~al.}(2014)\citenamefont
  {Ridgers}, \citenamefont {Kirk}, \citenamefont {Duclous}, \citenamefont
  {Blackburn}, \citenamefont {Brady}, \citenamefont {Bennett}, \citenamefont
  {Arber},\ and\ \citenamefont {Bell}}]{ridgers14}%
  \BibitemOpen
  \bibfield  {author} {\bibinfo {author} {\bibfnamefont {C.}~\bibnamefont
  {Ridgers}}, \bibinfo {author} {\bibfnamefont {J.~G.}\ \bibnamefont {Kirk}},
  \bibinfo {author} {\bibfnamefont {R.}~\bibnamefont {Duclous}}, \bibinfo
  {author} {\bibfnamefont {T.}~\bibnamefont {Blackburn}}, \bibinfo {author}
  {\bibfnamefont {C.}~\bibnamefont {Brady}}, \bibinfo {author} {\bibfnamefont
  {K.}~\bibnamefont {Bennett}}, \bibinfo {author} {\bibfnamefont
  {T.}~\bibnamefont {Arber}}, \ and\ \bibinfo {author} {\bibfnamefont
  {A.}~\bibnamefont {Bell}},\ }\href@noop {} {\bibfield  {journal} {\bibinfo
  {journal} {Journal of Computational Physics}\ }\textbf {\bibinfo {volume}
  {260}},\ \bibinfo {pages} {273} (\bibinfo {year} {2014})}\BibitemShut
  {NoStop}%
\bibitem [{\citenamefont {Gonoskov}\ \emph {et~al.}(2015)\citenamefont
  {Gonoskov}, \citenamefont {Bastrakov}, \citenamefont {Efimenko},
  \citenamefont {Ilderton}, \citenamefont {Marklund}, \citenamefont {Meyerov},
  \citenamefont {Muraviev}, \citenamefont {Sergeev}, \citenamefont {Surmin},\
  and\ \citenamefont {Wallin}}]{PRE2015Gonoskov}%
  \BibitemOpen
  \bibfield  {author} {\bibinfo {author} {\bibfnamefont {A.}~\bibnamefont
  {Gonoskov}}, \bibinfo {author} {\bibfnamefont {S.}~\bibnamefont {Bastrakov}},
  \bibinfo {author} {\bibfnamefont {E.}~\bibnamefont {Efimenko}}, \bibinfo
  {author} {\bibfnamefont {A.}~\bibnamefont {Ilderton}}, \bibinfo {author}
  {\bibfnamefont {M.}~\bibnamefont {Marklund}}, \bibinfo {author}
  {\bibfnamefont {I.}~\bibnamefont {Meyerov}}, \bibinfo {author} {\bibfnamefont
  {A.}~\bibnamefont {Muraviev}}, \bibinfo {author} {\bibfnamefont
  {A.}~\bibnamefont {Sergeev}}, \bibinfo {author} {\bibfnamefont
  {I.}~\bibnamefont {Surmin}}, \ and\ \bibinfo {author} {\bibfnamefont
  {E.}~\bibnamefont {Wallin}},\ }\href {\doibase 10.1103/PhysRevE.92.023305}
  {\bibfield  {journal} {\bibinfo  {journal} {Phys. Rev. E}\ }\textbf {\bibinfo
  {volume} {92}},\ \bibinfo {pages} {023305} (\bibinfo {year}
  {2015})}\BibitemShut {NoStop}%
\bibitem [{\citenamefont {Nikishov}\ and\ \citenamefont
  {Ritus}(1964)}]{nikishov64}%
  \BibitemOpen
  \bibfield  {author} {\bibinfo {author} {\bibfnamefont {A.~I.}\ \bibnamefont
  {Nikishov}}\ and\ \bibinfo {author} {\bibfnamefont {V.~I.}\ \bibnamefont
  {Ritus}},\ }\href@noop {} {\bibfield  {journal} {\bibinfo  {journal} {Sov.
  Phys. JETP}\ }\textbf {\bibinfo {volume} {19}},\ \bibinfo {pages} {529}
  (\bibinfo {year} {1964})}\BibitemShut {NoStop}%
\bibitem [{\citenamefont {Brown}\ and\ \citenamefont
  {Kibble}(1964)}]{kibble64}%
  \BibitemOpen
  \bibfield  {author} {\bibinfo {author} {\bibfnamefont {L.~S.}\ \bibnamefont
  {Brown}}\ and\ \bibinfo {author} {\bibfnamefont {T.~W.~B.}\ \bibnamefont
  {Kibble}},\ }\href@noop {} {\bibfield  {journal} {\bibinfo  {journal} {Phys.
  Rep.}\ }\textbf {\bibinfo {volume} {133}},\ \bibinfo {pages} {A705} (\bibinfo
  {year} {1964})}\BibitemShut {NoStop}%
\bibitem [{\citenamefont {Ritus}(1985)}]{ritus85}%
  \BibitemOpen
  \bibfield  {author} {\bibinfo {author} {\bibfnamefont {V.~I.}\ \bibnamefont
  {Ritus}},\ }\href@noop {} {\bibfield  {journal} {\bibinfo  {journal} {J.
  Russ. Laser Res.}\ }\textbf {\bibinfo {volume} {6}},\ \bibinfo {pages} {497}
  (\bibinfo {year} {1985})}\BibitemShut {NoStop}%
\bibitem [{\citenamefont {King}(2015)}]{king15b}%
  \BibitemOpen
  \bibfield  {author} {\bibinfo {author} {\bibfnamefont {B.}~\bibnamefont
  {King}},\ }\href@noop {} {\bibfield  {journal} {\bibinfo  {journal} {Phys.
  Rev. A}\ }\textbf {\bibinfo {volume} {91}},\ \bibinfo {pages} {033415}
  (\bibinfo {year} {2015})}\BibitemShut {NoStop}%
\bibitem [{\citenamefont {Seipt}\ and\ \citenamefont
  {King}(2020)}]{Seipt:2020diz}%
  \BibitemOpen
  \bibfield  {author} {\bibinfo {author} {\bibfnamefont {D.}~\bibnamefont
  {Seipt}}\ and\ \bibinfo {author} {\bibfnamefont {B.}~\bibnamefont {King}},\
  }\href@noop {} {\  (\bibinfo {year} {2020})},\ \Eprint
  {http://arxiv.org/abs/2007.11837} {arXiv:2007.11837 [physics.plasm-ph]}
  \BibitemShut {NoStop}%
\bibitem [{\citenamefont {Boca}\ and\ \citenamefont
  {Florescu}(2009)}]{Boca:2009zz}%
  \BibitemOpen
  \bibfield  {author} {\bibinfo {author} {\bibfnamefont {M.}~\bibnamefont
  {Boca}}\ and\ \bibinfo {author} {\bibfnamefont {V.}~\bibnamefont
  {Florescu}},\ }\href {\doibase 10.1103/PhysRevA.80.053403} {\bibfield
  {journal} {\bibinfo  {journal} {Phys. Rev. A}\ }\textbf {\bibinfo {volume}
  {80}},\ \bibinfo {pages} {053403} (\bibinfo {year} {2009})}\BibitemShut
  {NoStop}%
\bibitem [{\citenamefont {Krajewska}\ and\ \citenamefont
  {Kaminski}(2013)}]{krajewska13}%
  \BibitemOpen
  \bibfield  {author} {\bibinfo {author} {\bibfnamefont {K.}~\bibnamefont
  {Krajewska}}\ and\ \bibinfo {author} {\bibfnamefont {J.}~\bibnamefont
  {Kaminski}},\ }\href@noop {} {\bibfield  {journal} {\bibinfo  {journal}
  {Laser and Particle Beams}\ }\textbf {\bibinfo {volume} {31}},\ \bibinfo
  {pages} {503} (\bibinfo {year} {2013})}\BibitemShut {NoStop}%
\bibitem [{\citenamefont {Krajewska}\ and\ \citenamefont
  {Kami\ifmmode~\acute{n}\else \'{n}\fi{}ski}(2014)}]{PRA052117}%
  \BibitemOpen
  \bibfield  {author} {\bibinfo {author} {\bibfnamefont {K.}~\bibnamefont
  {Krajewska}}\ and\ \bibinfo {author} {\bibfnamefont {J.~Z.}\ \bibnamefont
  {Kami\ifmmode~\acute{n}\else \'{n}\fi{}ski}},\ }\href {\doibase
  10.1103/PhysRevA.90.052117} {\bibfield  {journal} {\bibinfo  {journal} {Phys.
  Rev. A}\ }\textbf {\bibinfo {volume} {90}},\ \bibinfo {pages} {052117}
  (\bibinfo {year} {2014})}\BibitemShut {NoStop}%
\bibitem [{\citenamefont {Jansen}\ \emph {et~al.}(2016)\citenamefont {Jansen},
  \citenamefont {Kami\ifmmode~\acute{n}\else \'{n}\fi{}ski}, \citenamefont
  {Krajewska},\ and\ \citenamefont {M\"uller}}]{PRD013010}%
  \BibitemOpen
  \bibfield  {author} {\bibinfo {author} {\bibfnamefont {M.~J.~A.}\
  \bibnamefont {Jansen}}, \bibinfo {author} {\bibfnamefont {J.~Z.}\
  \bibnamefont {Kami\ifmmode~\acute{n}\else \'{n}\fi{}ski}}, \bibinfo {author}
  {\bibfnamefont {K.}~\bibnamefont {Krajewska}}, \ and\ \bibinfo {author}
  {\bibfnamefont {C.}~\bibnamefont {M\"uller}},\ }\href {\doibase
  10.1103/PhysRevD.94.013010} {\bibfield  {journal} {\bibinfo  {journal} {Phys.
  Rev. D}\ }\textbf {\bibinfo {volume} {94}},\ \bibinfo {pages} {013010}
  (\bibinfo {year} {2016})}\BibitemShut {NoStop}%
\bibitem [{\citenamefont {Wistisen}\ and\ \citenamefont
  {Di~Piazza}(2019)}]{PRD116001}%
  \BibitemOpen
  \bibfield  {author} {\bibinfo {author} {\bibfnamefont {T.~N.}\ \bibnamefont
  {Wistisen}}\ and\ \bibinfo {author} {\bibfnamefont {A.}~\bibnamefont
  {Di~Piazza}},\ }\href {\doibase 10.1103/PhysRevD.100.116001} {\bibfield
  {journal} {\bibinfo  {journal} {Phys. Rev. D}\ }\textbf {\bibinfo {volume}
  {100}},\ \bibinfo {pages} {116001} (\bibinfo {year} {2019})}\BibitemShut
  {NoStop}%
\bibitem [{\citenamefont {Seipt}\ \emph {et~al.}(2018)\citenamefont {Seipt},
  \citenamefont {Del~Sorbo}, \citenamefont {Ridgers},\ and\ \citenamefont
  {Thomas}}]{seipt18}%
  \BibitemOpen
  \bibfield  {author} {\bibinfo {author} {\bibfnamefont {D.}~\bibnamefont
  {Seipt}}, \bibinfo {author} {\bibfnamefont {D.}~\bibnamefont {Del~Sorbo}},
  \bibinfo {author} {\bibfnamefont {C.}~\bibnamefont {Ridgers}}, \ and\
  \bibinfo {author} {\bibfnamefont {A.~R.}\ \bibnamefont {Thomas}},\ }\href
  {\doibase 10.1103/PhysRevA.98.023417} {\bibfield  {journal} {\bibinfo
  {journal} {Phys. Rev. A}\ }\textbf {\bibinfo {volume} {98}},\ \bibinfo
  {pages} {023417} (\bibinfo {year} {2018})},\ \Eprint
  {http://arxiv.org/abs/1805.02027} {arXiv:1805.02027 [hep-ph]} \BibitemShut
  {NoStop}%
\bibitem [{\citenamefont {Del~Sorbo}\ \emph {et~al.}(2017)\citenamefont
  {Del~Sorbo}, \citenamefont {Seipt}, \citenamefont {Blackburn}, \citenamefont
  {Thomas}, \citenamefont {Murphy}, \citenamefont {Kirk},\ and\ \citenamefont
  {Ridgers}}]{DelSorbo:2017fod}%
  \BibitemOpen
  \bibfield  {author} {\bibinfo {author} {\bibfnamefont {D.}~\bibnamefont
  {Del~Sorbo}}, \bibinfo {author} {\bibfnamefont {D.}~\bibnamefont {Seipt}},
  \bibinfo {author} {\bibfnamefont {T.~G.}\ \bibnamefont {Blackburn}}, \bibinfo
  {author} {\bibfnamefont {A.~G.~R.}\ \bibnamefont {Thomas}}, \bibinfo {author}
  {\bibfnamefont {C.~D.}\ \bibnamefont {Murphy}}, \bibinfo {author}
  {\bibfnamefont {J.~G.}\ \bibnamefont {Kirk}}, \ and\ \bibinfo {author}
  {\bibfnamefont {C.~P.}\ \bibnamefont {Ridgers}},\ }\href {\doibase
  10.1103/PhysRevA.96.043407} {\bibfield  {journal} {\bibinfo  {journal} {Phys.
  Rev. A}\ }\textbf {\bibinfo {volume} {96}},\ \bibinfo {pages} {043407}
  (\bibinfo {year} {2017})}\BibitemShut {NoStop}%
\bibitem [{\citenamefont {Sorbo}\ \emph {et~al.}(2018)\citenamefont {Sorbo},
  \citenamefont {Seipt}, \citenamefont {Thomas},\ and\ \citenamefont
  {Ridgers}}]{Del_Sorbo_2018}%
  \BibitemOpen
  \bibfield  {author} {\bibinfo {author} {\bibfnamefont {D.~D.}\ \bibnamefont
  {Sorbo}}, \bibinfo {author} {\bibfnamefont {D.}~\bibnamefont {Seipt}},
  \bibinfo {author} {\bibfnamefont {A.~G.~R.}\ \bibnamefont {Thomas}}, \ and\
  \bibinfo {author} {\bibfnamefont {C.~P.}\ \bibnamefont {Ridgers}},\ }\href
  {\doibase 10.1088/1361-6587/aab979} {\bibfield  {journal} {\bibinfo
  {journal} {Plasma Physics and Controlled Fusion}\ }\textbf {\bibinfo {volume}
  {60}},\ \bibinfo {pages} {064003} (\bibinfo {year} {2018})}\BibitemShut
  {NoStop}%
\bibitem [{\citenamefont {Seipt}\ \emph {et~al.}(2019)\citenamefont {Seipt},
  \citenamefont {Del~Sorbo}, \citenamefont {Ridgers},\ and\ \citenamefont
  {Thomas}}]{Seipt:2019ddd}%
  \BibitemOpen
  \bibfield  {author} {\bibinfo {author} {\bibfnamefont {D.}~\bibnamefont
  {Seipt}}, \bibinfo {author} {\bibfnamefont {D.}~\bibnamefont {Del~Sorbo}},
  \bibinfo {author} {\bibfnamefont {C.~P.}\ \bibnamefont {Ridgers}}, \ and\
  \bibinfo {author} {\bibfnamefont {A.~G.~R.}\ \bibnamefont {Thomas}},\ }\href
  {\doibase 10.1103/PhysRevA.100.061402} {\bibfield  {journal} {\bibinfo
  {journal} {Phys. Rev. A}\ }\textbf {\bibinfo {volume} {100}},\ \bibinfo
  {pages} {061402} (\bibinfo {year} {2019})}\BibitemShut {NoStop}%
\bibitem [{\citenamefont {Ba\u{\i}er}\ \emph {et~al.}(1976)\citenamefont
  {Ba\u{\i}er}, \citenamefont {Katkov}, \citenamefont {Mil'shte\u{\i}n},\ and\
  \citenamefont {Strakhovenko}}]{baier76}%
  \BibitemOpen
  \bibfield  {author} {\bibinfo {author} {\bibfnamefont {V.~N.}\ \bibnamefont
  {Ba\u{\i}er}}, \bibinfo {author} {\bibfnamefont {V.~M.}\ \bibnamefont
  {Katkov}}, \bibinfo {author} {\bibfnamefont {A.~I.}\ \bibnamefont
  {Mil'shte\u{\i}n}}, \ and\ \bibinfo {author} {\bibfnamefont {V.~M.}\
  \bibnamefont {Strakhovenko}},\ }\href@noop {} {\bibfield  {journal} {\bibinfo
   {journal} {Sov. Phys. JETP}\ }\textbf {\bibinfo {volume} {42}},\ \bibinfo
  {pages} {400} (\bibinfo {year} {1976})}\BibitemShut {NoStop}%
\bibitem [{\citenamefont {Meuren}\ and\ \citenamefont
  {Di~Piazza}(2011)}]{meuren11}%
  \BibitemOpen
  \bibfield  {author} {\bibinfo {author} {\bibfnamefont {S.}~\bibnamefont
  {Meuren}}\ and\ \bibinfo {author} {\bibfnamefont {A.}~\bibnamefont
  {Di~Piazza}},\ }\href@noop {} {\bibfield  {journal} {\bibinfo  {journal}
  {Phys. Rev. Lett.}\ }\textbf {\bibinfo {volume} {107}},\ \bibinfo {pages}
  {260401} (\bibinfo {year} {2011})}\BibitemShut {NoStop}%
\bibitem [{\citenamefont {Lavelle}\ and\ \citenamefont
  {McMullan}(2019)}]{LAVELLE2019135021}%
  \BibitemOpen
  \bibfield  {author} {\bibinfo {author} {\bibfnamefont {M.}~\bibnamefont
  {Lavelle}}\ and\ \bibinfo {author} {\bibfnamefont {D.}~\bibnamefont
  {McMullan}},\ }\href {\doibase
  https://doi.org/10.1016/j.physletb.2019.135021} {\bibfield  {journal}
  {\bibinfo  {journal} {Physics Letters B}\ }\textbf {\bibinfo {volume}
  {798}},\ \bibinfo {pages} {135021} (\bibinfo {year} {2019})}\BibitemShut
  {NoStop}%
\bibitem [{\citenamefont {Ritus}(1970)}]{Ritus1}%
  \BibitemOpen
  \bibfield  {author} {\bibinfo {author} {\bibfnamefont {V.}~\bibnamefont
  {Ritus}},\ }\href@noop {} {\bibfield  {journal} {\bibinfo  {journal} {Sov.
  Phys. JETP}\ }\textbf {\bibinfo {volume} {30}},\ \bibinfo {pages} {1113}
  (\bibinfo {year} {1970})}\BibitemShut {NoStop}%
\bibitem [{\citenamefont {Narozhnyi}(1979)}]{Narozhnyi:1979at}%
  \BibitemOpen
  \bibfield  {author} {\bibinfo {author} {\bibfnamefont {N.}~\bibnamefont
  {Narozhnyi}},\ }\href {\doibase 10.1103/PhysRevD.20.1313} {\bibfield
  {journal} {\bibinfo  {journal} {Phys. Rev. D}\ }\textbf {\bibinfo {volume}
  {20}},\ \bibinfo {pages} {1313} (\bibinfo {year} {1979})}\BibitemShut
  {NoStop}%
\bibitem [{\citenamefont {Narozhnyi}(1980)}]{Narozhnyi:1980dc}%
  \BibitemOpen
  \bibfield  {author} {\bibinfo {author} {\bibfnamefont {N.}~\bibnamefont
  {Narozhnyi}},\ }\href {\doibase 10.1103/PhysRevD.21.1176} {\bibfield
  {journal} {\bibinfo  {journal} {Phys. Rev. D}\ }\textbf {\bibinfo {volume}
  {21}},\ \bibinfo {pages} {1176} (\bibinfo {year} {1980})}\BibitemShut
  {NoStop}%
\bibitem [{\citenamefont {Fedotov}(2017)}]{Fedotov:2016afw}%
  \BibitemOpen
  \bibfield  {author} {\bibinfo {author} {\bibfnamefont {A.}~\bibnamefont
  {Fedotov}},\ }\href {\doibase 10.1088/1742-6596/826/1/012027} {\bibfield
  {journal} {\bibinfo  {journal} {J. Phys. Conf. Ser.}\ }\textbf {\bibinfo
  {volume} {826}},\ \bibinfo {pages} {012027} (\bibinfo {year} {2017})},\
  \Eprint {http://arxiv.org/abs/1608.02261} {arXiv:1608.02261 [hep-ph]}
  \BibitemShut {NoStop}%
\bibitem [{\citenamefont {Podszus}\ and\ \citenamefont
  {Di~Piazza}(2019)}]{Podszus:2018hnz}%
  \BibitemOpen
  \bibfield  {author} {\bibinfo {author} {\bibfnamefont {T.}~\bibnamefont
  {Podszus}}\ and\ \bibinfo {author} {\bibfnamefont {A.}~\bibnamefont
  {Di~Piazza}},\ }\href {\doibase 10.1103/PhysRevD.99.076004} {\bibfield
  {journal} {\bibinfo  {journal} {Phys. Rev. D}\ }\textbf {\bibinfo {volume}
  {99}},\ \bibinfo {pages} {076004} (\bibinfo {year} {2019})}\BibitemShut
  {NoStop}%
\bibitem [{\citenamefont {Ilderton}(2019{\natexlab{a}})}]{Ilderton:2019kqp}%
  \BibitemOpen
  \bibfield  {author} {\bibinfo {author} {\bibfnamefont {A.}~\bibnamefont
  {Ilderton}},\ }\href {\doibase 10.1103/PhysRevD.99.085002} {\bibfield
  {journal} {\bibinfo  {journal} {Phys. Rev. D}\ }\textbf {\bibinfo {volume}
  {99}},\ \bibinfo {pages} {085002} (\bibinfo {year}
  {2019}{\natexlab{a}})}\BibitemShut {NoStop}%
\bibitem [{\citenamefont {Ilderton}(2019{\natexlab{b}})}]{Ilderton:2019vot}%
  \BibitemOpen
  \bibfield  {author} {\bibinfo {author} {\bibfnamefont {A.}~\bibnamefont
  {Ilderton}},\ }\href {\doibase 10.1103/PhysRevD.100.125018} {\bibfield
  {journal} {\bibinfo  {journal} {Phys. Rev.}\ }\textbf {\bibinfo {volume}
  {D100}},\ \bibinfo {pages} {125018} (\bibinfo {year} {2019}{\natexlab{b}})},\
  \Eprint {http://arxiv.org/abs/1909.02484} {arXiv:1909.02484 [hep-ph]}
  \BibitemShut {NoStop}%
\bibitem [{\citenamefont {Mironov}\ \emph {et~al.}(2020)\citenamefont
  {Mironov}, \citenamefont {Meuren},\ and\ \citenamefont
  {Fedotov}}]{Mironov:2020gbi}%
  \BibitemOpen
  \bibfield  {author} {\bibinfo {author} {\bibfnamefont {A.}~\bibnamefont
  {Mironov}}, \bibinfo {author} {\bibfnamefont {S.}~\bibnamefont {Meuren}}, \
  and\ \bibinfo {author} {\bibfnamefont {A.}~\bibnamefont {Fedotov}},\
  }\href@noop {} {\  (\bibinfo {year} {2020})},\ \Eprint
  {http://arxiv.org/abs/2003.06909} {arXiv:2003.06909 [hep-th]} \BibitemShut
  {NoStop}%
\bibitem [{\citenamefont {Yakimenko}\ \emph {et~al.}(2019)\citenamefont
  {Yakimenko} \emph {et~al.}}]{Yakimenko:2018kih}%
  \BibitemOpen
  \bibfield  {author} {\bibinfo {author} {\bibfnamefont {V.}~\bibnamefont
  {Yakimenko}} \emph {et~al.},\ }\href {\doibase
  10.1103/PhysRevLett.122.190404} {\bibfield  {journal} {\bibinfo  {journal}
  {Phys. Rev. Lett.}\ }\textbf {\bibinfo {volume} {122}},\ \bibinfo {pages}
  {190404} (\bibinfo {year} {2019})},\ \Eprint
  {http://arxiv.org/abs/1807.09271} {arXiv:1807.09271 [physics.plasm-ph]}
  \BibitemShut {NoStop}%
\bibitem [{\citenamefont {Baumann}\ \emph {et~al.}(2019)\citenamefont
  {Baumann}, \citenamefont {Nerush}, \citenamefont {Pukhov},\ and\
  \citenamefont {Kostyukov}}]{Baumann:2018ovl}%
  \BibitemOpen
  \bibfield  {author} {\bibinfo {author} {\bibfnamefont {C.}~\bibnamefont
  {Baumann}}, \bibinfo {author} {\bibfnamefont {E.}~\bibnamefont {Nerush}},
  \bibinfo {author} {\bibfnamefont {A.}~\bibnamefont {Pukhov}}, \ and\ \bibinfo
  {author} {\bibfnamefont {I.}~\bibnamefont {Kostyukov}},\ }\href {\doibase
  10.1038/s41598-019-45582-5} {\bibfield  {journal} {\bibinfo  {journal} {Sci.
  Rep.}\ }\textbf {\bibinfo {volume} {9}},\ \bibinfo {pages} {9407} (\bibinfo
  {year} {2019})},\ \Eprint {http://arxiv.org/abs/1811.03990} {arXiv:1811.03990
  [physics.plasm-ph]} \BibitemShut {NoStop}%
\bibitem [{\citenamefont {Blackburn}\ \emph {et~al.}(2019)\citenamefont
  {Blackburn}, \citenamefont {Ilderton}, \citenamefont {Marklund},\ and\
  \citenamefont {Ridgers}}]{Blackburn:2018tsn}%
  \BibitemOpen
  \bibfield  {author} {\bibinfo {author} {\bibfnamefont {T.}~\bibnamefont
  {Blackburn}}, \bibinfo {author} {\bibfnamefont {A.}~\bibnamefont {Ilderton}},
  \bibinfo {author} {\bibfnamefont {M.}~\bibnamefont {Marklund}}, \ and\
  \bibinfo {author} {\bibfnamefont {C.}~\bibnamefont {Ridgers}},\ }\href
  {\doibase 10.1088/1367-2630/ab1e0d} {\bibfield  {journal} {\bibinfo
  {journal} {New J. Phys.}\ }\textbf {\bibinfo {volume} {21}},\ \bibinfo
  {pages} {053040} (\bibinfo {year} {2019})},\ \Eprint
  {http://arxiv.org/abs/1807.03730} {arXiv:1807.03730 [physics.plasm-ph]}
  \BibitemShut {NoStop}%
\bibitem [{\citenamefont {Di~Piazza}\ \emph {et~al.}(2020)\citenamefont
  {Di~Piazza}, \citenamefont {Wistisen}, \citenamefont {Tamburini},\ and\
  \citenamefont {Uggerh{\o}j}}]{DiPiazza:2019vwb}%
  \BibitemOpen
  \bibfield  {author} {\bibinfo {author} {\bibfnamefont {A.}~\bibnamefont
  {Di~Piazza}}, \bibinfo {author} {\bibfnamefont {T.}~\bibnamefont {Wistisen}},
  \bibinfo {author} {\bibfnamefont {M.}~\bibnamefont {Tamburini}}, \ and\
  \bibinfo {author} {\bibfnamefont {U.}~\bibnamefont {Uggerh{\o}j}},\ }\href
  {\doibase 10.1103/PhysRevLett.124.044801} {\bibfield  {journal} {\bibinfo
  {journal} {Phys. Rev. Lett.}\ }\textbf {\bibinfo {volume} {124}},\ \bibinfo
  {pages} {044801} (\bibinfo {year} {2020})},\ \Eprint
  {http://arxiv.org/abs/1911.04749} {arXiv:1911.04749 [hep-ph]} \BibitemShut
  {NoStop}%
\bibitem [{\citenamefont {Li}\ \emph {et~al.}(2019)\citenamefont {Li},
  \citenamefont {Shaisultanov}, \citenamefont {Hatsagortsyan}, \citenamefont
  {Wan}, \citenamefont {Keitel},\ and\ \citenamefont {Li}}]{Li:2019PRL}%
  \BibitemOpen
  \bibfield  {author} {\bibinfo {author} {\bibfnamefont {Y.-F.}\ \bibnamefont
  {Li}}, \bibinfo {author} {\bibfnamefont {R.}~\bibnamefont {Shaisultanov}},
  \bibinfo {author} {\bibfnamefont {K.~Z.}\ \bibnamefont {Hatsagortsyan}},
  \bibinfo {author} {\bibfnamefont {F.}~\bibnamefont {Wan}}, \bibinfo {author}
  {\bibfnamefont {C.~H.}\ \bibnamefont {Keitel}}, \ and\ \bibinfo {author}
  {\bibfnamefont {J.-X.}\ \bibnamefont {Li}},\ }\href {\doibase
  10.1103/PhysRevLett.122.154801} {\bibfield  {journal} {\bibinfo  {journal}
  {Phys. Rev. Lett.}\ }\textbf {\bibinfo {volume} {122}},\ \bibinfo {pages}
  {154801} (\bibinfo {year} {2019})}\BibitemShut {NoStop}%
\bibitem [{\citenamefont {Wan}\ \emph {et~al.}(2020)\citenamefont {Wan},
  \citenamefont {Shaisultanov}, \citenamefont {Li}, \citenamefont
  {Hatsagortsyan}, \citenamefont {Keitel},\ and\ \citenamefont
  {Li}}]{WAN2020135120}%
  \BibitemOpen
  \bibfield  {author} {\bibinfo {author} {\bibfnamefont {F.}~\bibnamefont
  {Wan}}, \bibinfo {author} {\bibfnamefont {R.}~\bibnamefont {Shaisultanov}},
  \bibinfo {author} {\bibfnamefont {Y.-F.}\ \bibnamefont {Li}}, \bibinfo
  {author} {\bibfnamefont {K.~Z.}\ \bibnamefont {Hatsagortsyan}}, \bibinfo
  {author} {\bibfnamefont {C.~H.}\ \bibnamefont {Keitel}}, \ and\ \bibinfo
  {author} {\bibfnamefont {J.-X.}\ \bibnamefont {Li}},\ }\href {\doibase
  https://doi.org/10.1016/j.physletb.2019.135120} {\bibfield  {journal}
  {\bibinfo  {journal} {Physics Letters B}\ }\textbf {\bibinfo {volume}
  {800}},\ \bibinfo {pages} {135120} (\bibinfo {year} {2020})}\BibitemShut
  {NoStop}%
\bibitem [{\citenamefont {Thomas}\ \emph {et~al.}(2020)\citenamefont {Thomas},
  \citenamefont {H\"utzen}, \citenamefont {Lehrach}, \citenamefont {Pukhov},
  \citenamefont {Ji}, \citenamefont {Wu}, \citenamefont {Geng},\ and\
  \citenamefont {B\"uscher}}]{PRAB064401}%
  \BibitemOpen
  \bibfield  {author} {\bibinfo {author} {\bibfnamefont {J.}~\bibnamefont
  {Thomas}}, \bibinfo {author} {\bibfnamefont {A.}~\bibnamefont {H\"utzen}},
  \bibinfo {author} {\bibfnamefont {A.}~\bibnamefont {Lehrach}}, \bibinfo
  {author} {\bibfnamefont {A.}~\bibnamefont {Pukhov}}, \bibinfo {author}
  {\bibfnamefont {L.}~\bibnamefont {Ji}}, \bibinfo {author} {\bibfnamefont
  {Y.}~\bibnamefont {Wu}}, \bibinfo {author} {\bibfnamefont {X.}~\bibnamefont
  {Geng}}, \ and\ \bibinfo {author} {\bibfnamefont {M.}~\bibnamefont
  {B\"uscher}},\ }\href {\doibase 10.1103/PhysRevAccelBeams.23.064401}
  {\bibfield  {journal} {\bibinfo  {journal} {Phys. Rev. Accel. Beams}\
  }\textbf {\bibinfo {volume} {23}},\ \bibinfo {pages} {064401} (\bibinfo
  {year} {2020})}\BibitemShut {NoStop}%
\bibitem [{\citenamefont {Baier}(1972)}]{baier72b}%
  \BibitemOpen
  \bibfield  {author} {\bibinfo {author} {\bibfnamefont {V.~N.}\ \bibnamefont
  {Baier}},\ }\href@noop {} {\bibfield  {journal} {\bibinfo  {journal} {Soviet
  Phys. Usp.}\ }\textbf {\bibinfo {volume} {14}},\ \bibinfo {pages} {695}
  (\bibinfo {year} {1972})}\BibitemShut {NoStop}%
\bibitem [{\citenamefont {Leader}\ and\ \citenamefont
  {Lorc\'e}(2014)}]{Leader:2013jra}%
  \BibitemOpen
  \bibfield  {author} {\bibinfo {author} {\bibfnamefont {E.}~\bibnamefont
  {Leader}}\ and\ \bibinfo {author} {\bibfnamefont {C.}~\bibnamefont
  {Lorc\'e}},\ }\href {\doibase 10.1016/j.physrep.2014.02.010} {\bibfield
  {journal} {\bibinfo  {journal} {Phys. Rept.}\ }\textbf {\bibinfo {volume}
  {541}},\ \bibinfo {pages} {163} (\bibinfo {year} {2014})},\ \Eprint
  {http://arxiv.org/abs/1309.4235} {arXiv:1309.4235 [hep-ph]} \BibitemShut
  {NoStop}%
\bibitem [{\citenamefont {Dirac}(1949)}]{Dirac:1949cp}%
  \BibitemOpen
  \bibfield  {author} {\bibinfo {author} {\bibfnamefont {P.~A.}\ \bibnamefont
  {Dirac}},\ }\href {\doibase 10.1103/RevModPhys.21.392} {\bibfield  {journal}
  {\bibinfo  {journal} {Rev. Mod. Phys.}\ }\textbf {\bibinfo {volume} {21}},\
  \bibinfo {pages} {392} (\bibinfo {year} {1949})}\BibitemShut {NoStop}%
\bibitem [{\citenamefont {Brodsky}\ \emph {et~al.}(1998)\citenamefont
  {Brodsky}, \citenamefont {Pauli},\ and\ \citenamefont
  {Pinsky}}]{Brodsky:1997de}%
  \BibitemOpen
  \bibfield  {author} {\bibinfo {author} {\bibfnamefont {S.~J.}\ \bibnamefont
  {Brodsky}}, \bibinfo {author} {\bibfnamefont {H.-C.}\ \bibnamefont {Pauli}},
  \ and\ \bibinfo {author} {\bibfnamefont {S.~S.}\ \bibnamefont {Pinsky}},\
  }\href {\doibase 10.1016/S0370-1573(97)00089-6} {\bibfield  {journal}
  {\bibinfo  {journal} {Phys. Rept.}\ }\textbf {\bibinfo {volume} {301}},\
  \bibinfo {pages} {299} (\bibinfo {year} {1998})},\ \Eprint
  {http://arxiv.org/abs/hep-ph/9705477} {arXiv:hep-ph/9705477 [hep-ph]}
  \BibitemShut {NoStop}%
\bibitem [{\citenamefont {Heinzl}(2001)}]{Heinzl:2000ht}%
  \BibitemOpen
  \bibfield  {author} {\bibinfo {author} {\bibfnamefont {T.}~\bibnamefont
  {Heinzl}},\ }\href {\doibase 10.1007/3-540-45114-5_2} {\bibfield  {journal}
  {\bibinfo  {journal} {Lect. Notes Phys.}\ }\textbf {\bibinfo {volume}
  {572}},\ \bibinfo {pages} {55} (\bibinfo {year} {2001})},\ \bibinfo {note}
  {{Methods of quantization. Proceedings, 39. Internationale
  Universitätswochen für Kern- und Teilchenphysik, IUKT 39: Schladming,
  Austria, February 26-March 4, 2000}},\ \Eprint
  {http://arxiv.org/abs/hep-th/0008096} {arXiv:hep-th/0008096 [hep-th]}
  \BibitemShut {NoStop}%
\bibitem [{\citenamefont {Bakker}\ \emph {et~al.}(2014)\citenamefont {Bakker}
  \emph {et~al.}}]{Bakker:2013cea}%
  \BibitemOpen
  \bibfield  {author} {\bibinfo {author} {\bibfnamefont {B.}~\bibnamefont
  {Bakker}} \emph {et~al.},\ }\href {\doibase
  10.1016/j.nuclphysbps.2014.05.004} {\bibfield  {journal} {\bibinfo  {journal}
  {Nucl. Phys. B Proc. Suppl.}\ }\textbf {\bibinfo {volume} {251-252}},\
  \bibinfo {pages} {165} (\bibinfo {year} {2014})},\ \Eprint
  {http://arxiv.org/abs/1309.6333} {arXiv:1309.6333 [hep-ph]} \BibitemShut
  {NoStop}%
\bibitem [{\citenamefont {Dinu}\ \emph {et~al.}(2012)\citenamefont {Dinu},
  \citenamefont {Heinzl},\ and\ \citenamefont {Ilderton}}]{Dinu:2012tj}%
  \BibitemOpen
  \bibfield  {author} {\bibinfo {author} {\bibfnamefont {V.}~\bibnamefont
  {Dinu}}, \bibinfo {author} {\bibfnamefont {T.}~\bibnamefont {Heinzl}}, \ and\
  \bibinfo {author} {\bibfnamefont {A.}~\bibnamefont {Ilderton}},\ }\href
  {\doibase 10.1103/PhysRevD.86.085037} {\bibfield  {journal} {\bibinfo
  {journal} {Phys. Rev. D}\ }\textbf {\bibinfo {volume} {86}},\ \bibinfo
  {pages} {085037} (\bibinfo {year} {2012})},\ \Eprint
  {http://arxiv.org/abs/1206.3957} {arXiv:1206.3957 [hep-ph]} \BibitemShut
  {NoStop}%
\bibitem [{\citenamefont {Bargmann}\ \emph {et~al.}(1959)\citenamefont
  {Bargmann}, \citenamefont {Michel},\ and\ \citenamefont
  {Telegdi}}]{bargmann59}%
  \BibitemOpen
  \bibfield  {author} {\bibinfo {author} {\bibfnamefont {V.}~\bibnamefont
  {Bargmann}}, \bibinfo {author} {\bibfnamefont {L.}~\bibnamefont {Michel}}, \
  and\ \bibinfo {author} {\bibfnamefont {V.~L.}\ \bibnamefont {Telegdi}},\
  }\href {\doibase 10.1103/PhysRevLett.2.435} {\bibfield  {journal} {\bibinfo
  {journal} {Phys. Rev. Lett.}\ }\textbf {\bibinfo {volume} {2}},\ \bibinfo
  {pages} {435} (\bibinfo {year} {1959})}\BibitemShut {NoStop}%
\bibitem [{\citenamefont {Jacob}\ and\ \citenamefont {Wick}(1959)}]{jacob59}%
  \BibitemOpen
  \bibfield  {author} {\bibinfo {author} {\bibfnamefont {M.}~\bibnamefont
  {Jacob}}\ and\ \bibinfo {author} {\bibfnamefont {G.}~\bibnamefont {Wick}},\
  }\href {\doibase https://doi.org/10.1016/0003-4916(59)90051-X} {\bibfield
  {journal} {\bibinfo  {journal} {Annals of Physics}\ }\textbf {\bibinfo
  {volume} {7}},\ \bibinfo {pages} {404 } (\bibinfo {year} {1959})}\BibitemShut
  {NoStop}%
\bibitem [{\citenamefont {Chiu}\ and\ \citenamefont
  {Brodsky}(2017)}]{Chiu:2017ycx}%
  \BibitemOpen
  \bibfield  {author} {\bibinfo {author} {\bibfnamefont {K.~Y.-J.}\
  \bibnamefont {Chiu}}\ and\ \bibinfo {author} {\bibfnamefont {S.~J.}\
  \bibnamefont {Brodsky}},\ }\href {\doibase 10.1103/PhysRevD.95.065035}
  {\bibfield  {journal} {\bibinfo  {journal} {Phys. Rev. D}\ }\textbf {\bibinfo
  {volume} {95}},\ \bibinfo {pages} {065035} (\bibinfo {year} {2017})},\
  \Eprint {http://arxiv.org/abs/1702.01127} {arXiv:1702.01127 [hep-th]}
  \BibitemShut {NoStop}%
\bibitem [{\citenamefont {Itzykson}\ and\ \citenamefont
  {Zuber}(1980)}]{Itzykson:1980rh}%
  \BibitemOpen
  \bibfield  {author} {\bibinfo {author} {\bibfnamefont {C.}~\bibnamefont
  {Itzykson}}\ and\ \bibinfo {author} {\bibfnamefont {J.}~\bibnamefont
  {Zuber}},\ }\href@noop {} {\emph {\bibinfo {title} {{Quantum Field
  Theory}}}}\ (\bibinfo  {publisher} {McGraw-Hill},\ \bibinfo {address} {New
  York},\ \bibinfo {year} {1980})\BibitemShut {NoStop}%
\bibitem [{\citenamefont {Volkov}(1935)}]{volkov35}%
  \BibitemOpen
  \bibfield  {author} {\bibinfo {author} {\bibfnamefont {D.~M.}\ \bibnamefont
  {Volkov}},\ }\href@noop {} {\bibfield  {journal} {\bibinfo  {journal} {Z.
  Phys.}\ }\textbf {\bibinfo {volume} {94}},\ \bibinfo {pages} {250} (\bibinfo
  {year} {1935})}\BibitemShut {NoStop}%
\bibitem [{\citenamefont {Aleksandrov}\ \emph {et~al.}(2020)\citenamefont
  {Aleksandrov}, \citenamefont {Tumakov}, \citenamefont {Kudlis}, \citenamefont
  {Shabaev},\ and\ \citenamefont {Rosanov}}]{Aleksandrov2020}%
  \BibitemOpen
  \bibfield  {author} {\bibinfo {author} {\bibfnamefont {I.~A.}\ \bibnamefont
  {Aleksandrov}}, \bibinfo {author} {\bibfnamefont {D.~A.}\ \bibnamefont
  {Tumakov}}, \bibinfo {author} {\bibfnamefont {A.}~\bibnamefont {Kudlis}},
  \bibinfo {author} {\bibfnamefont {V.~M.}\ \bibnamefont {Shabaev}}, \ and\
  \bibinfo {author} {\bibfnamefont {N.~N.}\ \bibnamefont {Rosanov}},\ }\href
  {\doibase 10.1103/PhysRevA.102.023102} {\bibfield  {journal} {\bibinfo
  {journal} {Phys. Rev. A}\ }\textbf {\bibinfo {volume} {102}},\ \bibinfo
  {pages} {023102} (\bibinfo {year} {2020})}\BibitemShut {NoStop}%
\bibitem [{\citenamefont {Dinu}(2013)}]{Dinu:PRA052101}%
  \BibitemOpen
  \bibfield  {author} {\bibinfo {author} {\bibfnamefont {V.}~\bibnamefont
  {Dinu}},\ }\href {\doibase 10.1103/PhysRevA.87.052101} {\bibfield  {journal}
  {\bibinfo  {journal} {Phys. Rev. A}\ }\textbf {\bibinfo {volume} {87}},\
  \bibinfo {pages} {052101} (\bibinfo {year} {2013})}\BibitemShut {NoStop}%
\bibitem [{\citenamefont {Seipt}(2017)}]{seipt2017volkov}%
  \BibitemOpen
  \bibfield  {author} {\bibinfo {author} {\bibfnamefont {D.}~\bibnamefont
  {Seipt}},\ }\href@noop {} {\bibfield  {journal} {\bibinfo  {journal} {arXiv
  preprint arXiv:1701.03692}\ } (\bibinfo {year} {2017})}\BibitemShut {NoStop}%
\bibitem [{\citenamefont {Ilderton}\ and\ \citenamefont
  {Torgrimsson}(2013)}]{ilderton2013scattering}%
  \BibitemOpen
  \bibfield  {author} {\bibinfo {author} {\bibfnamefont {A.}~\bibnamefont
  {Ilderton}}\ and\ \bibinfo {author} {\bibfnamefont {G.}~\bibnamefont
  {Torgrimsson}},\ }\href@noop {} {\bibfield  {journal} {\bibinfo  {journal}
  {Physical Review D}\ }\textbf {\bibinfo {volume} {87}},\ \bibinfo {pages}
  {085040} (\bibinfo {year} {2013})}\BibitemShut {NoStop}%
\bibitem [{\citenamefont {Bieri}\ and\ \citenamefont
  {Garfinkle}(2013)}]{Bieri:2013hqa}%
  \BibitemOpen
  \bibfield  {author} {\bibinfo {author} {\bibfnamefont {L.}~\bibnamefont
  {Bieri}}\ and\ \bibinfo {author} {\bibfnamefont {D.}~\bibnamefont
  {Garfinkle}},\ }\href {\doibase 10.1088/0264-9381/30/19/195009} {\bibfield
  {journal} {\bibinfo  {journal} {Class. Quant. Grav.}\ }\textbf {\bibinfo
  {volume} {30}},\ \bibinfo {pages} {195009} (\bibinfo {year} {2013})},\
  \Eprint {http://arxiv.org/abs/1307.5098} {arXiv:1307.5098 [gr-qc]}
  \BibitemShut {NoStop}%
\bibitem [{\citenamefont {Di~Piazza}\ \emph {et~al.}(2019)\citenamefont
  {Di~Piazza}, \citenamefont {Tamburini}, \citenamefont {Meuren},\ and\
  \citenamefont {Keitel}}]{dipiazza18}%
  \BibitemOpen
  \bibfield  {author} {\bibinfo {author} {\bibfnamefont {A.}~\bibnamefont
  {Di~Piazza}}, \bibinfo {author} {\bibfnamefont {M.}~\bibnamefont
  {Tamburini}}, \bibinfo {author} {\bibfnamefont {S.}~\bibnamefont {Meuren}}, \
  and\ \bibinfo {author} {\bibfnamefont {C.~H.}\ \bibnamefont {Keitel}},\
  }\href {\doibase 10.1103/PhysRevA.99.022125} {\bibfield  {journal} {\bibinfo
  {journal} {Phys. Rev. A}\ }\textbf {\bibinfo {volume} {99}},\ \bibinfo
  {pages} {022125} (\bibinfo {year} {2019})}\BibitemShut {NoStop}%
\bibitem [{\citenamefont {Ilderton}\ \emph {et~al.}(2019)\citenamefont
  {Ilderton}, \citenamefont {King},\ and\ \citenamefont
  {Seipt}}]{PRA2019Ilderton}%
  \BibitemOpen
  \bibfield  {author} {\bibinfo {author} {\bibfnamefont {A.}~\bibnamefont
  {Ilderton}}, \bibinfo {author} {\bibfnamefont {B.}~\bibnamefont {King}}, \
  and\ \bibinfo {author} {\bibfnamefont {D.}~\bibnamefont {Seipt}},\ }\href
  {\doibase 10.1103/PhysRevA.99.042121} {\bibfield  {journal} {\bibinfo
  {journal} {Phys. Rev. A}\ }\textbf {\bibinfo {volume} {99}},\ \bibinfo
  {pages} {042121} (\bibinfo {year} {2019})}\BibitemShut {NoStop}%
\bibitem [{\citenamefont {Olver}\ \emph {et~al.}(2010)\citenamefont {Olver},
  \citenamefont {Lozier}, \citenamefont {Boisvert},\ and\ \citenamefont
  {Clark}}]{olver2010nist}%
  \BibitemOpen
  \bibfield  {author} {\bibinfo {author} {\bibfnamefont {F.~W.}\ \bibnamefont
  {Olver}}, \bibinfo {author} {\bibfnamefont {D.~W.}\ \bibnamefont {Lozier}},
  \bibinfo {author} {\bibfnamefont {R.~F.}\ \bibnamefont {Boisvert}}, \ and\
  \bibinfo {author} {\bibfnamefont {C.~W.}\ \bibnamefont {Clark}},\ }\href@noop
  {} {\emph {\bibinfo {title} {NIST handbook of mathematical functions}}}\
  (\bibinfo  {publisher} {Cambridge university press},\ \bibinfo {year}
  {2010})\BibitemShut {NoStop}%
\bibitem [{\citenamefont {Harvey}\ \emph {et~al.}(2015)\citenamefont {Harvey},
  \citenamefont {Ilderton},\ and\ \citenamefont {King}}]{harvey15}%
  \BibitemOpen
  \bibfield  {author} {\bibinfo {author} {\bibfnamefont {C.~N.}\ \bibnamefont
  {Harvey}}, \bibinfo {author} {\bibfnamefont {A.}~\bibnamefont {Ilderton}}, \
  and\ \bibinfo {author} {\bibfnamefont {B.}~\bibnamefont {King}},\ }\href
  {\doibase 10.1103/PhysRevA.91.013822} {\bibfield  {journal} {\bibinfo
  {journal} {Phys. Rev. A}\ }\textbf {\bibinfo {volume} {91}},\ \bibinfo
  {pages} {013822} (\bibinfo {year} {2015})}\BibitemShut {NoStop}%
\bibitem [{\citenamefont {Di~Piazza}\ \emph {et~al.}(2018)\citenamefont
  {Di~Piazza}, \citenamefont {Tamburini}, \citenamefont {Meuren},\ and\
  \citenamefont {Keitel}}]{meuren17}%
  \BibitemOpen
  \bibfield  {author} {\bibinfo {author} {\bibfnamefont {A.}~\bibnamefont
  {Di~Piazza}}, \bibinfo {author} {\bibfnamefont {M.}~\bibnamefont
  {Tamburini}}, \bibinfo {author} {\bibfnamefont {S.}~\bibnamefont {Meuren}}, \
  and\ \bibinfo {author} {\bibfnamefont {C.~H.}\ \bibnamefont {Keitel}},\
  }\href {\doibase 10.1103/PhysRevA.98.012134} {\bibfield  {journal} {\bibinfo
  {journal} {Phys. Rev.}\ }\textbf {\bibinfo {volume} {A98}},\ \bibinfo {pages}
  {012134} (\bibinfo {year} {2018})},\ \Eprint
  {http://arxiv.org/abs/1708.08276} {arXiv:1708.08276 [hep-ph]} \BibitemShut
  {NoStop}%
\bibitem [{\citenamefont {Heinzl}\ \emph {et~al.}(2020)\citenamefont {Heinzl},
  \citenamefont {King},\ and\ \citenamefont {MacLeod}}]{heinzl2020locally}%
  \BibitemOpen
  \bibfield  {author} {\bibinfo {author} {\bibfnamefont {T.}~\bibnamefont
  {Heinzl}}, \bibinfo {author} {\bibfnamefont {B.}~\bibnamefont {King}}, \ and\
  \bibinfo {author} {\bibfnamefont {A.}~\bibnamefont {MacLeod}},\ }\href@noop
  {} {\bibfield  {journal} {\bibinfo  {journal} {arXiv preprint
  arXiv:2004.13035}\ } (\bibinfo {year} {2020})}\BibitemShut {NoStop}%
\bibitem [{\citenamefont {Salamin}\ and\ \citenamefont
  {Keitel}(2002)}]{YousefPRL095005}%
  \BibitemOpen
  \bibfield  {author} {\bibinfo {author} {\bibfnamefont {Y.~I.}\ \bibnamefont
  {Salamin}}\ and\ \bibinfo {author} {\bibfnamefont {C.~H.}\ \bibnamefont
  {Keitel}},\ }\href {\doibase 10.1103/PhysRevLett.88.095005} {\bibfield
  {journal} {\bibinfo  {journal} {Phys. Rev. Lett.}\ }\textbf {\bibinfo
  {volume} {88}},\ \bibinfo {pages} {095005} (\bibinfo {year}
  {2002})}\BibitemShut {NoStop}%
\end{thebibliography}
\providecommand{\noopsort}[1]{}
\end{document}